\documentclass[twocolumn,pra,secnumarabic,amssymb,amsmath,nobibnotes,aps,showpacs]{revtex4-1}

\usepackage{graphics}
\usepackage{graphicx}
\usepackage{braket}
\usepackage[usenames,dvipsnames]{color}

\begin{document}

\title{Feshbach resonances in ultracold $^{85}$Rb}

\author{Caroline L. Blackley, C. Ruth Le Sueur, Jeremy M. Hutson }
\affiliation{Joint Quantum Centre (JQC) Durham/Newcastle, Department of Chemistry, Durham University, South Road, Durham, DH1 3LE, United Kingdom}

\author{Daniel J. McCarron, Michael P. K\"oppinger, Hung-Wen Cho, Daniel L. Jenkin, and Simon L. Cornish}
\affiliation{Joint Quantum Centre (JQC) Durham/Newcastle, Department of
Physics, Durham University, South Road, Durham DH1 3LE, United Kingdom}

\begin{abstract}
We present 17 experimentally confirmed Feshbach resonances in optically trapped
$^{85}$Rb. Seven of the resonances are in the ground-state channel
$(f,m_f)=(2,+2)+(2,+2)$, and nine are in the excited-state channel
$(2,-2)+(2,-2)$. We find a wide resonance at high field in each of the two
channels, offering new possibilities for the formation of larger $^{85}$Rb
condensates and studies of few-body physics. A detailed coupled-channels
analysis is presented to characterize the resonances, and also provides an
understanding of the inelastic losses observed in the excited-state channel. In
addition we have confirmed the existence of one narrow resonance in a
$(2,+2)+(3,+3)$ spin mixture.
\end{abstract}
\date{\today}

\maketitle

%\pacs{}

\section{Introduction}

The creation of ultracold molecules is currently of great interest. They offer
a wide range of applications including: studies of few-body quantum physics,
high precision spectroscopy, quantum simulators for many-body phenomena and
controlled chemistry \cite{Carr:NJPintro:2009,Krems:IRPC:2005}. Ultracold
molecules have a far richer substructure than atoms and so molecular
condensates with tunable interactions offer unique levels of control over
collision properties \cite{Chin:RMP:2010}. One route to ultracold molecules is
through the association of two ultracold atoms into a weakly bound molecule
\cite{Damski:2003}. The energy of a bound molecular state is tuned
adiabatically through an avoided crossing with the energy of the separated
atomic states \cite{Chin:RMP:2010}, forming a weakly bound molecule. The
molecules can then be transferred into their ro-vibrational ground state by
stimulated Raman adiabatic passage (STIRAP). This method has been used
effectively in several systems to create ultracold molecules \cite{Ni:KRb:2008,
Lang:ground:2008, Danzl:ground:2010}.

$^{85}$Rb is a promising species for ultracold atomic gas experiments, though
it has often been overlooked due to the challenges of forming a Bose-Einstein
condensate (BEC) \cite{Burke:1998, Cornish:2000}. Our recent work shows the
benefits of $^{85}$Rb for RbCs production \cite{Cho:RbCs:2012}. However, for
these experiments a full understanding of the scattering behavior of $^{85}$Rb
is required. Most previous work on $^{85}$Rb has focused on the wide resonance
near 155~G in the $(f,m_f)=(2,-2)+(2,-2)$ channel \cite{Vogels:1997}. This
resonance is suitable for experiments that require precise tuning of the
scattering length and has been used extensively in studies of condensate
collapse \cite{Cornish:2000, Roberts:collapse:2001, Donley:2001, Altin:2011},
the formation of bright matter wave solitons \cite{Cornish:2006}, and few-body
physics \cite{Wild:2012}. Further work using $^{85}$Rb includes spectroscopic
studies of photo-association \cite{Tsai:1997,Courteille:1998} and measurements
of inelastic collision rates \cite{Roberts:2000, Roberts:elast:2001}, molecular
binding energies \cite{Claussen:2003}, molecule formation \cite{Donley:2002,
Thompson:magres:2005, Kohler:2005, Hodby:2005, Brown:2006} and Efimov states
\cite{Stoll:2005, Altin:2011, Wild:2012}. Despite extensive work in this region
of the $(2,-2)+(2,-2)$ channel, there appears to have been little theoretical
or experimental work on the ground state, or at higher field.

In this paper we reveal the rich Feshbach structure of $^{85}$Rb. We use
coupled-channels calculations to predict Feshbach resonances in both the
 $(2,-2)+(2,-2)$ channel (designated ee), and $(2,+2)+(2,+2)$ channel
(designated aa) and confirm 16 of them experimentally. In addition we identify
a resonance in the mixed spin channel $(2,+2)+(3,+3)$. The structure of the
paper is as follow: Section II describes the theory and calculations; Section
III describes the experimental setup and methodology; Section IV describes the
results, including an outlook on future research prospects.

\begin{figure*}[!]
\includegraphics[width=2\columnwidth]{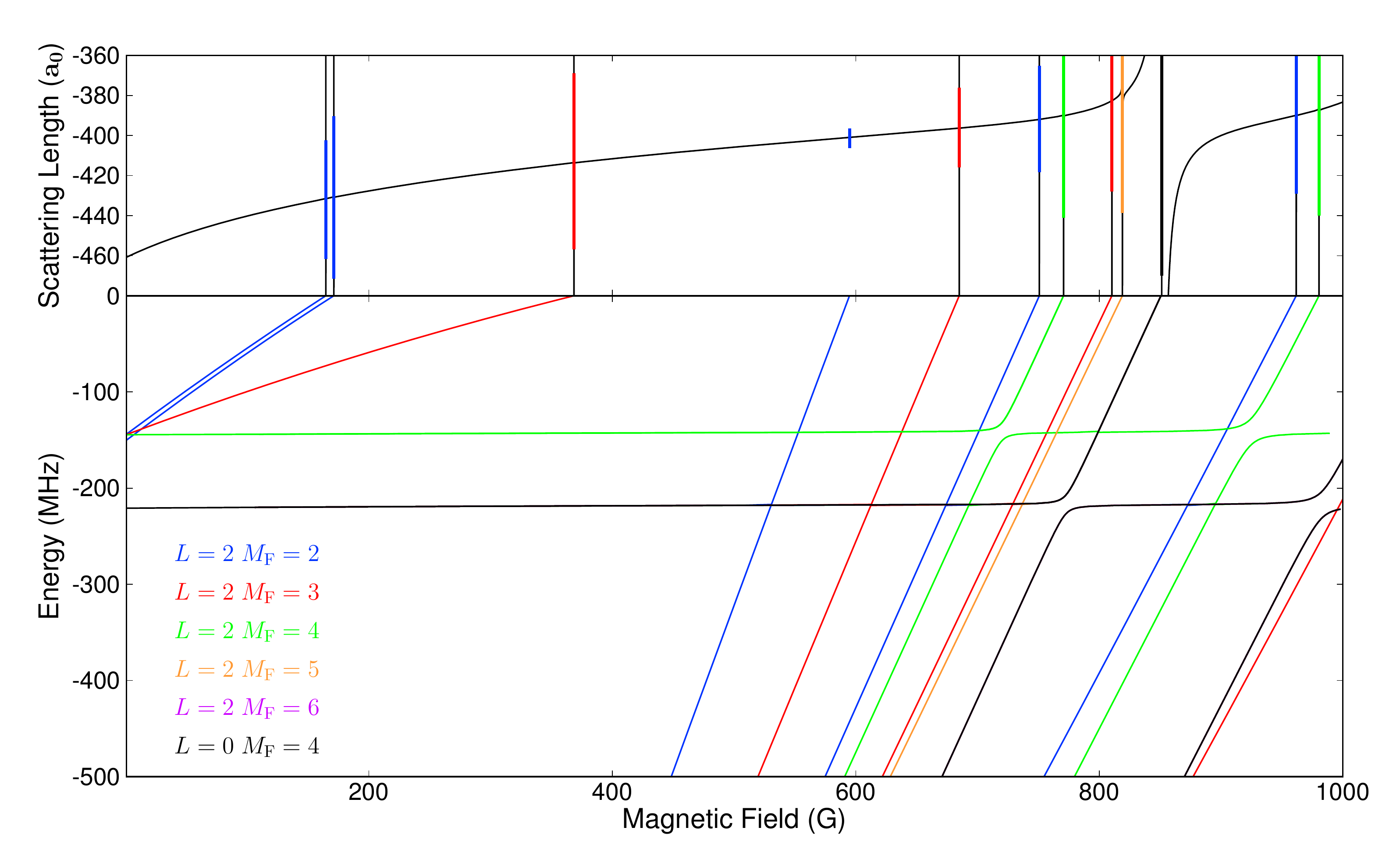}
\caption{(Color online) Upper panel: The calculated s-wave scattering length in
the aa channel of $^{85}$Rb$_2$, with resonances marked with lines whose color
depends on their $M_F$ value (see legend); the length of each line is
proportional to the logarithm of the width of the resonance. Lower panel: The
energies of weakly bound molecular states, relative to the aa threshold,
$(2,+2)+(2,+2)$~channel. All calculations in this figure are for
$M_{\rm{tot}}=4$, corresponding to s-wave scattering in the aa channel.}
\label{fig:aa_scattering}
\end{figure*}

\section{Theory}
The collision Hamiltonian for a pair of alkali-metal atoms is
\begin{equation}
\frac{\hbar^2}{2\mu}\bigg[-r^{-1}\frac{d^2}{dr^2}r
+\frac{\hat{L}^2}{r^2}\bigg]+\hat{H}_{\rm{1}}+\hat{H}_{\rm{2}}+\hat{V}(r),
\end{equation}
where $r$ is the internuclear distance, $\mu$ is the reduced mass, $\hat{L}$ is
the rotational angular momentum operator and $\hat{V}$ is the interaction
operator. $\hat H_1$ and $\hat H_2$ are the monomer Hamiltonians of the free
atoms,
\begin{equation}
\hat{H}_{i} = \zeta_{i}\hat{i}_{i}\cdot\hat{s}_{i}
+(g_{\rm{e}}\mu_{\rm{B}}\hat{s}_{iz}+g_{\rm{n}}\mu_{\rm{B}}\hat{i}_{iz})B,
\end{equation}
where $ \zeta_i$ is the hyperfine coupling constant of atom $i$,
$g_{\rm{e}}$ and $g_{\rm{n}}$ are the electron and nuclear g-factors, $\hat{s}$
and $\hat{i}$ are the electron and nuclear spin operators and $B$ is the
magnetic field.

The calculations in the present paper are carried out in two different basis
sets: a fully decoupled basis set
$$\ket{s_{\rm{Rb}}m_{s\rm{Rb}}}\ket{i_{\rm{Rb}}m_{i\rm{Rb}}}\ket{s_{\rm{Rb}}m_{s\rm{Rb}}}\ket{i_{\rm{Rb}}m_{i\rm{Rb}}}\ket{LM_{L}},$$
and a partly coupled basis set
$$\ket{f_a,m_{f,a}}\ket{f_b,m_{f,b}}\ket{F,M_F}\ket{L,m_L}.$$
The two basis sets give identical bound-state energies and scattering
properties, but different views of the bound-state wavefunctions. In both cases
the basis sets are symmetrized to take account of identical particle symmetry.
The resulting coupled equations are diagonal in the total projection number
$M_{\rm tot }= M_F + M_L$, where $M_F = m_{f,a}+m_{f,b} =
m_{s{\rm{Rb_a}}}+m_{i{\rm{Rb_a}}}+m_{s{\rm{Rb_b}}}+m_{i{\rm{Rb_b}}}$. The basis
sets used include all functions with $L=0$ and $2$ for the required value of
$M_{\rm tot}$, which for s-wave scattering is equal to $m_{f,a}+m_{f,b}$ for
the incoming channel.

\begin{figure*}[!]
\includegraphics[width=2\columnwidth]{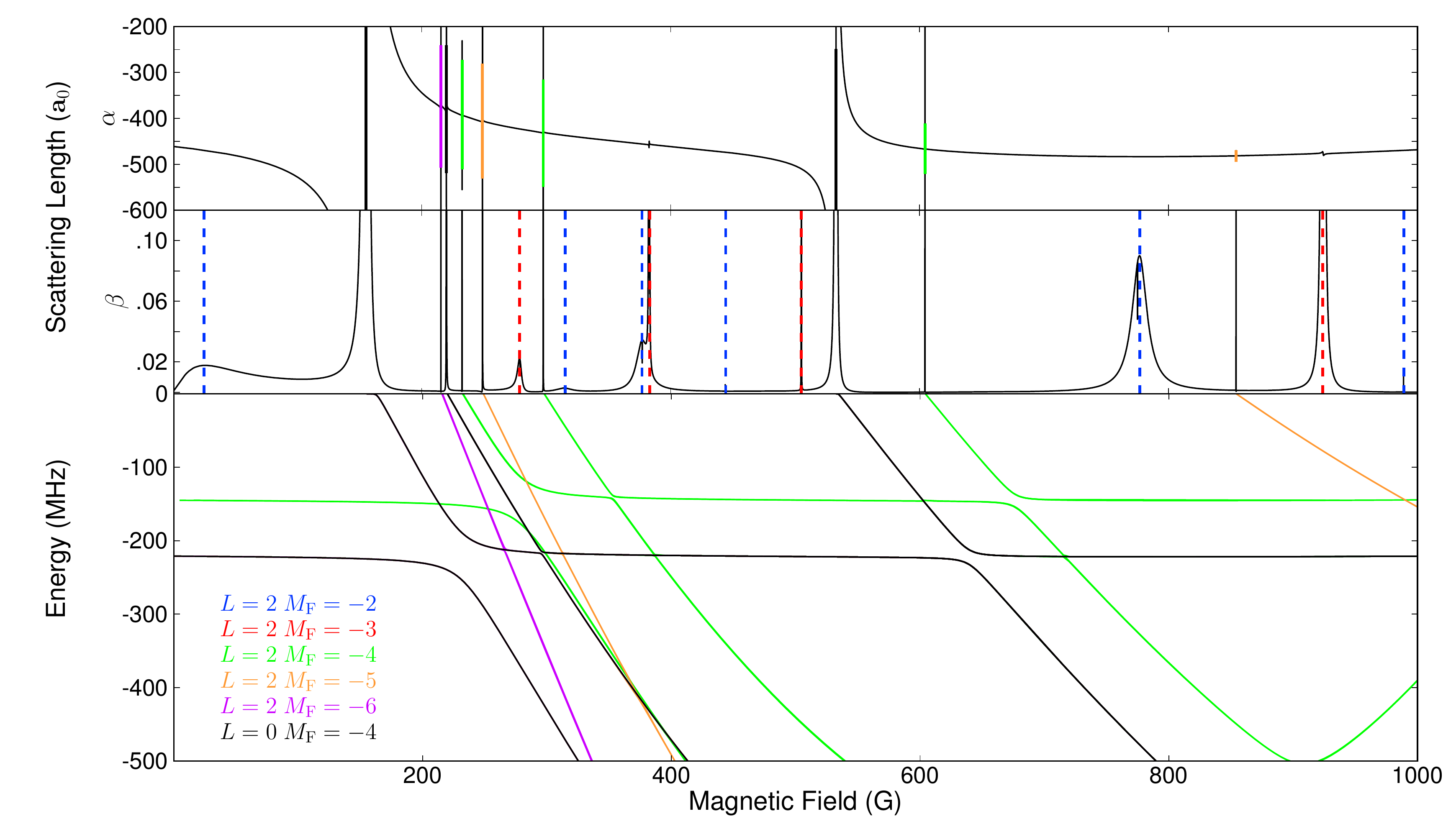}
\caption{(Color online) Upper panels: The real and imaginary parts of the
s-wave scattering length in the ee channel of $^{85}$Rb$_2$. Each resonance is
indicated by a colored vertical line that indicates its $M_F$ value (see
legend); for pole-like resonances, the length of the line is proportional to
the logarithm of the width of the resonance. Inelastically dominated resonances
are not always evident in $\alpha$ but appear as peaks in $\beta$ and are
indicated by dashed vertical lines. Lower panel: The energies of weakly bound
molecular states, relative to the ee threshold, $(2,-2)+(2,-2)$ channel. Only
states with no continuum interference ($M_F\le-4$) are shown in the bound-state
map, but all resonances are included in the scattering length. All calculations
in this figure are for $M_{\rm{tot}}=-4$, corresponding s-wave scattering in ee
channel.} \label{fig:ee_scattering}
\end{figure*}

The coupled-channel scattering calculations are performed using the MOLSCAT
program \cite{molscat:v14}, as modified to handle collisions in an external
field \cite{Gonzalez-Martinez:2007}.  Calculations are carried out with a
fixed-step log-derivative propagator \cite{Manolopoulos:1986} from 0.3~nm to
2.1~nm and a variable-step Airy propagator \cite{Alexander:1984} from 2.1~nm to
1,500~nm. The wavefunctions are matched to their long range solutions, the
Ricatti-Bessel functions, at 1,500~nm to find the S-matrix elements. The s-wave
($L=0$) scattering length $a(k)$ is then obtained from the identity $a(k) =
(ik)^{-1}(1-S_{00})/(1+S_{00})$ \cite{Hutson:res:2007}, where $S_{00}$ is the
diagonal S-matrix element in the incoming channel and $k$ is the wavevector.
The bound-state calculations use the BOUND \cite{Hutson:bound:1993} and FIELD
\cite{Hutson:field:2011} packages, which locate bound states using as described
in Ref \cite{Hutson:Cs2:2008}. BOUND and FIELD use propagator methods, without
radial basis sets. The calculations allow the assignment of quantum numbers to
the states responsible for resonances in the scattering length.

The scattering and bound-state calculations are carried out using the potential
curves and magnetic dipole coupling function from Ref.\ \cite{Strauss:2010}.
The potentials were obtained by fitting to spectroscopic data on both the
singlet \cite{Seto:2000} and triplet states of $^{87}$Rb$_2$ and the triplet
state of $^{85}$Rb$_2$, together with several Feshbach resonances in
$^{87}$Rb$_2$, $^{87}$Rb$^{85}$Rb and $^{85}$Rb$_2$. The singlet and triplet
scattering lengths for $^{85}$Rb on the potentials of ref.\ \cite{Strauss:2010}
are $a_S = 2735$~$a_0$ and $a_T =-386$~$a_0$ respectively.

The calculated s-wave scattering length for the aa channel is shown in the top
panel of Figure \ref{fig:aa_scattering} and the binding energies of the
near-threshold molecular states responsible for the resonances are shown in the
lower panel. The resonance positions are given in Table
\ref{table:85RbFeshbachResonances_aa}, along with their widths $\Delta$ as
defined by local fits to the standard formula $a(B) =
a_{\rm{bg}}\left[1-\Delta/(B-B_0)\right]$ \cite{Moerdijk:1995}, where $a_{\rm
bg}$ is the background scattering length, $\Delta$ is the width, and $B_0$ is
the position of the pole in the scattering length. Quantum numbers were
assigned by carrying out approximate calculations with either $M_F$ or $F$ and
$M_F$ restricted to specific values. For a homonuclear diatomic molecule, $F$
is a nearly good quantum number in the low-field region where the free-atom
energies vary linearly with $B$. Figure \ref{fig:aa_scattering} shows one wide
resonance near 851~G ($\Delta=1.2$~G) that offers attractive possibilities for
precise tuning of the scattering length, and many narrower resonances that may
be useful for molecule formation.

For an excited-state channel, where inelastic scattering can occur, the
scattering length $a(B)$ is complex, $a(B) = \alpha(B) - i\beta(B)$. The
two-body inelastic loss rate is proportional to $\beta(B)$. The upper panels of
Figure \ref{fig:ee_scattering} show the real and imaginary parts of $a(B)$ for
s-wave collisions in the ee channel. In this case the inelastic collisions
produce atoms in lower magnetic sublevels, with $m_{f,a}$ and/or $m_{f,b} >
-2$. The lower panel shows the corresponding molecular bound states for
$M_F=-4$, $-5$ and $-6$, obtained from calculations with $M_F$ fixed. We also
carried out calculations of the quasibound states with $M_F=-2$ and $-3$ near
the ee threshold in order to identify the states responsible for the remaining
resonances. These calculations use the FIELD program with propagation to
reduced distances around 100 nm in order to reduce interference from continuum
states.

In the presence of inelastic scattering, $a(B)$ does not show actual poles at
resonance \cite{Hutson:res:2007}. If the background inelastic scattering is
negligible, the real part $\alpha(B)$ shows an oscillation of amplitude $a_{\rm
res}$, while the imaginary part shows a peak of height $a_{\rm res}$. The
resonant scattering length $a_{\rm res}$ is determined by the {\em ratio} of
the couplings from the quasibound state responsible for the resonance to the
incoming and inelastic channels \cite{Hutson:res:2007}. If there is significant
background scattering, then there is a more complicated asymmetric lineshape
that may show a substantial dip in the inelastic scattering near resonance
\cite{Hutson:HeO2:2009}. Figure \ref{fig:ee_scattering} shows resonances of all
these different types: the resonances due to bound states with
$M_F=m_{f,a}+m_{f,b}=-4$, $-5$ and $-6$ are pole-like, with values of at least
$a_{\rm res} > 20$ $a_0$ and with most $a_{\rm res} > 1000$ $a_0$. These
resonances produce pronounced features in $\alpha(B)$ and sharp peaks in
$\beta(B)$, off scale in Figure \ref{fig:ee_scattering}. By contrast,
resonances due to states with $M_F=-2$ and $-3$ show much weaker features with
$a_{\rm res} < 15$ $a_0$ and some lower than 0.01 $a_0$. These are barely
perceptible in $\alpha(B)$ on the scale of Figure \ref{fig:ee_scattering} and
produce broader, weaker peaks in $\beta(B)$. The distinction occurs because all
the inelastic channels have $M_F>-4$: bound states with $M_F=-2$ and $-3$ are
generally more strongly coupled to inelastic channels with the same $M_F$ than
to the incoming channel with $M_F=-4$, whereas the reverse is true for bound
states with $M_F=-4$, $-5$ and $-6$. Many of the features show quite pronounced
asymmetry in the shape of the inelastic peaks. All of the resonances with
$a_{\rm res} > 1.0$ $a_0$ are listed in Table
\ref{table:85RbFeshbachResonances_ee} along with their widths and approximate
$a_{\rm res}$ values.

We have also investigated the scattering length for mixed spin channels with a
view to identify broad resonances suitable for manipulating interactions. Most
channels exhibit strong inelastic decay with measured trap lifetimes of
$\sim100$~ms. However, the $(2,+2)+(3,+3)$ channel is immune to inelastic spin
exchange collisions, resulting in trap lifetimes of $\sim5$~s. The scattering
length in the mixed spin channel, $(2,+2)+(3,+3)$ shows two pole-like
resonances at 818.8~G and 909.9 G, both with widths of $2$~mG, and $a_{\rm
res}$= 1600 $a_0$ and 800 $a_0$ respectively.

\section{Experiment}
The details of our apparatus and cooling scheme are presented in Refs.\
\cite{McCarron2011,Cho2011,Cho:RbCs:2012} and are only briefly recounted here.
Ultracold samples of $^{85}$Rb are collected in a magneto-optical trap before
being optically pumped into the $(2,-2)$ state and loaded into a magnetic
quadrupole trap. Forced RF evaporation cools the $^{85}$Rb atoms to 50~$\mu$K
where further efficient evaporation is impeded by Majorana losses
\cite{Lin2009}. The atoms are then transferred into a crossed dipole trap
derived from a single-mode 1550~nm, 30W fibre laser. When loading, the power in
each beam is set to 4~W, creating a trap 100~$\mu$K deep with radial and axial
trap frequencies of 455~Hz and 90~Hz respectively. After loading, when
performing Feshbach spectroscopy in the absolute internal ground state, the
$^{85}$Rb atoms are transferred into the $|2,+2\rangle$ state by RF adiabatic
passage \cite{Jenkin2011}. A vertical magnetic field gradient of 21.2~G/cm is
then applied, slightly below the 22.4~G/cm necessary to levitate $^{85}$Rb.  In
contrast, when working with the $(2,-2)$ state, no magnetic field
gradient is applied and the atoms are confined in a purely optical potential.

A typical experiment begins with \mbox{$6.0(3)\times10^{5}$} $^{85}$Rb atoms at
$10.2(1)~\mu$K confined in the dipole trap in either $(2,+2)$ or $(2,-2)$. To
perform Feshbach spectroscopy, the magnetic field is switched to a specific
value in the range 0 to 1000~G. Evaporative cooling is then performed by
reducing the dipole beam powers by a factor of 4 over 2~s. The atomic sample is
then held for 1~s in this final potential. Resonant absorption imaging is used
to probe the atoms after each experimental cycle. Feshbach resonances are
identified by examining the variation in the atom number and temperature as a
function of the magnetic field. The magnetic field is calibrated using
microwave spectroscopy between the hyperfine states of $^{85}$Rb. These
measurements reveal our field stability to be $0.1$~G for the range 0 to 400~G
and $0.5$~G for the range 400 to 1000~G.

To perform Feshbach spectroscopy on a spin mixture of $(2,+2)+(3,+3)$, a cloud
of $(2,+2)$ atoms is first cooled using the same evaporation sequence as above
at 22.6~G. To populate the $(3,+3)$ state a microwave pulse is applied for
250~ms at 3.0887~GHz producing a mixture containing $7(1)\times10^{4}$ atoms in
each of the $(2,+2)$ and $(3,+3)$ spin states. The magnetic field is then
switched to a value in the range 0 to 1000~G and held for 750~ms. Finally,
Stern-Gerlach spectroscopy and absorption imaging are used to probe both spin
states simultaneously.

\section{Experimental Results}
We have observed 7 resonances in the aa channel and 9 resonances in the ee
channel. The observed and predicted resonance positions and widths are listed
in Tables \ref{table:85RbFeshbachResonances_aa} and
\ref{table:85RbFeshbachResonances_ee} and show good agreement between
experiment and theory.  In the ground state, all the widest calculated
resonances are seen experimentally, with the exception of the two high-field
resonances where the experimental field is less stable. In the excited state,
all but two of the predicted pole-like resonances are seen, together with two
of the inelastically dominated features.

\begin{figure}[t]
\includegraphics[width=1\columnwidth]{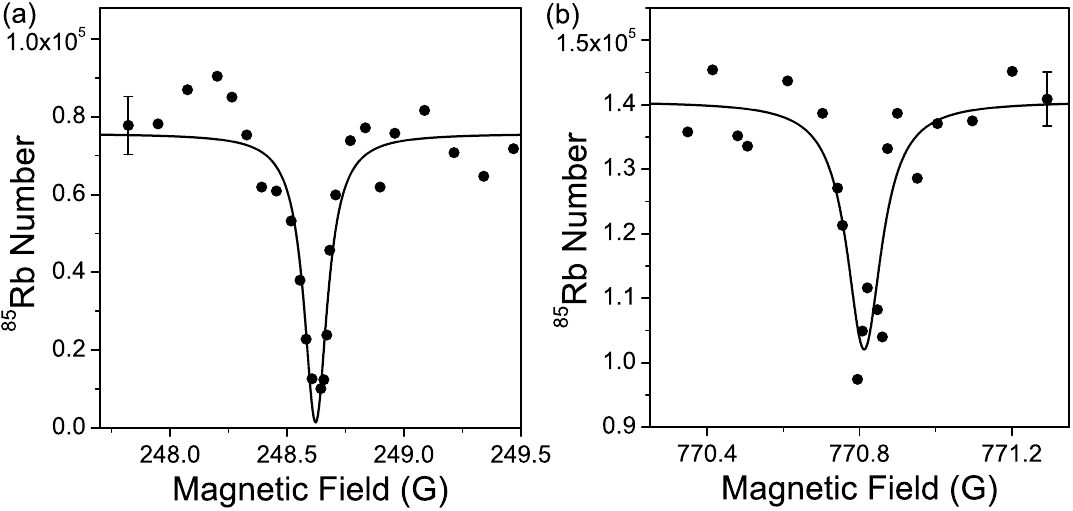}
\caption{Narrow resonances in $^{85}$Rb, with fitted width $\delta<0.2$~G,
observed as loss features in the atom number. (a) A resonance in the $(2,-2)$
state at 248.64(1)~G. (b) A resonance in the $(2,+2)$ state at 770.81(1)~G.
The error bars show the standard deviation for multiple control shots at specific
magnetic fields.} \label{fig:Narrow}
\end{figure}

\begin{figure}[t]
\includegraphics[width=1\columnwidth]{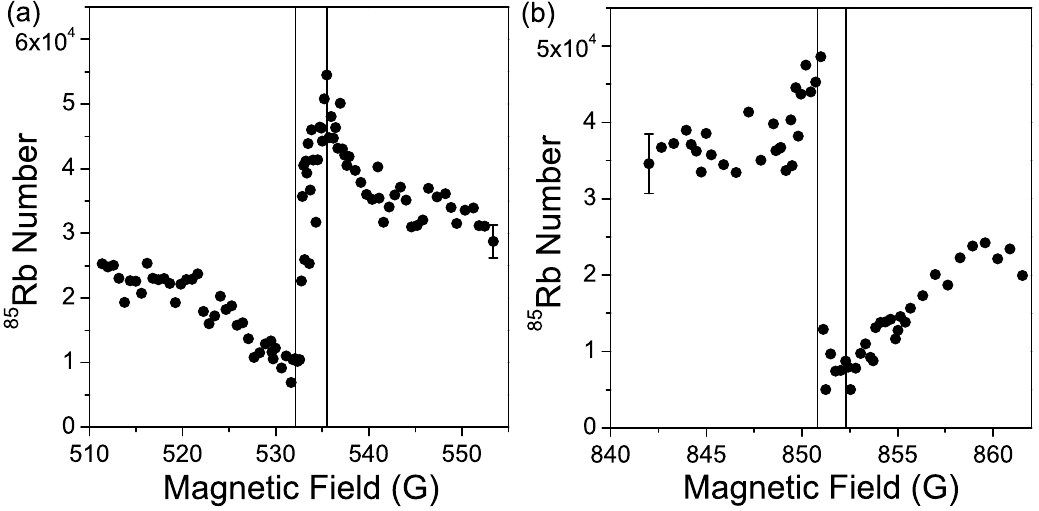}
\caption{Broad resonances in $^{85}$Rb, with width $\Delta>1$~G, observed as
features in the atom number. (a) A resonance in the $(2,-2)$ state at
532.3(3)~G. (b) A resonance in the $(2,+2)$ state at 852.3(3)~G. The
experimental widths are determined by the difference between the positions of
the minima and maxima in the atom number marked with solid lines in both plots.
The error bars show the standard deviation for multiple control shots at
specific magnetic fields.} \label{fig:Broad}
\end{figure}

Figure \ref{fig:Narrow} shows fine scans of the atom number for two of the
narrow resonances, one in each of the aa and ee channels. Such resonances
produce sharp drops in atom number. The experimental positions and widths of
these resonances are determined by fitting a Lorentzian, with width $\delta$,
to the data points. It should be noted that the experimental and theoretical
widths are entirely different quantities for narrow resonances, and should not
be compared.

\begin{figure}[t]
\includegraphics[width=1\columnwidth]{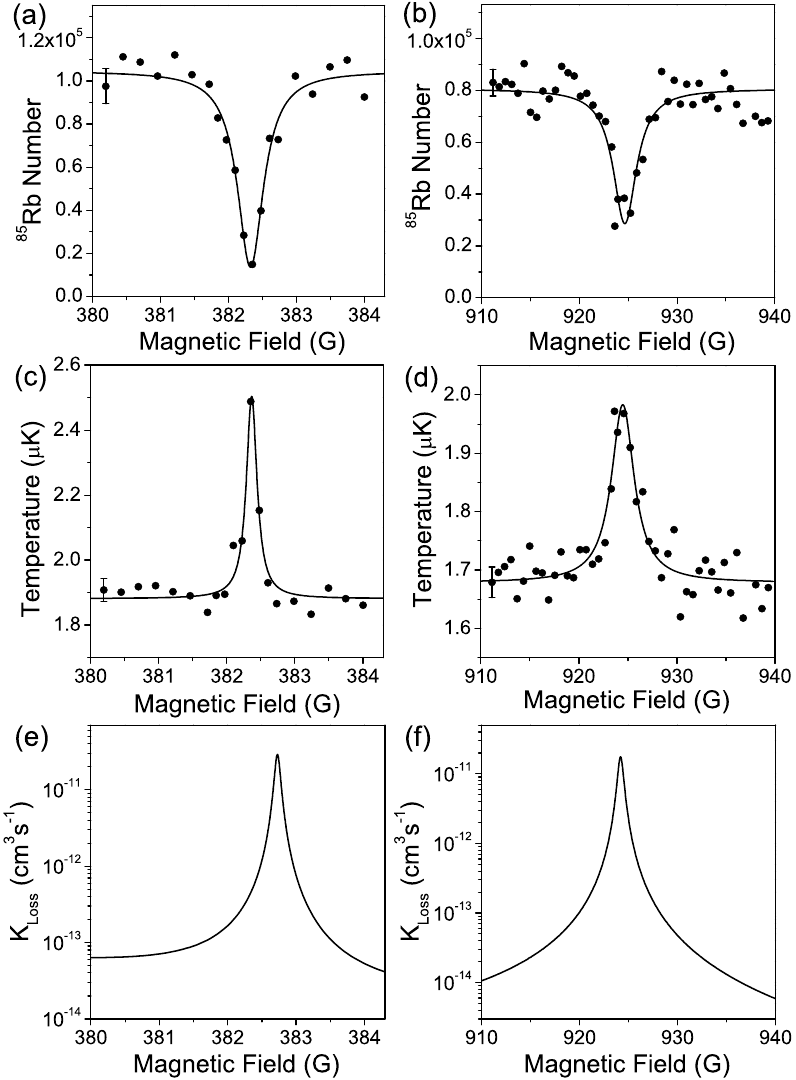}
\caption{The two inelastically dominated features observed in the $(2,-2)$
state. Figures (a) and (b) show the atom number while (c) and (d) show
temperature. Error bars show the standard deviation for multiple control shots
at specific magnetic fields. (e) and (f) show the calculated rate coefficient
for two-body loss $K_{\rm loss}$.} \label{fig:Inelastic}
\end{figure}

There are three resonances with widths greater than 1~G. Figure \ref{fig:Broad}
shows a fine scan across the resonances near 530~G in the ee channel, and 850~G
in the aa channel. In these cases the atom number shows both a peak and a
trough. The trough (loss maximum) again corresponds to the resonance position,
while the peak (loss minimum) occurs near the zero-crossing of the scattering
length. The three wide resonances are several orders of magnitude wider than
any of the other resonances seen and provide valuable control over the
scattering length. Note that our measurement of the position of the well-known
155~G resonance in the ee channel is not as accurate as the determination from
bound-state spectroscopy \cite{Claussen:2003}.

\begin{table}[!]
\begin{center}
\begin{tabular}{cccccccccc}
\hline
\hline
\multicolumn{10}{c} {Incoming s-wave (2,2)+(2,2) state} \\
\hline
 \multicolumn{2}{c} {Experiment}& & \multicolumn{7}{c} {Theory}\\
 \cline{1-2}
 \cline{4-10}
 $B_0$ & $\delta$ & & \multicolumn{4}{c} {Assignment} & $B_0$ & $\Delta$ & $a_{\rm{bg}}$\\
 \cline{4-7}
  (G) & (G) &  & $L$ & ($f_a$,$f_b$) & $F$ & $M_{F}$ & (G) & (mG) &  (bohr) \\
\hline
 164.74(1) & 0.08(2) &&2&(2,2)&4&2&164.7 & $-0.0006$ & $-432$ \\
 171.36(1) & 0.12(2) &&2&(2,2)&2&2&171.3 & $-0.02$ & $-431$ \\
 368.78(4) & 0.4(1)  &&2&(2,2)&4&3&368.5 & $-0.06$ & $-413$ \\
  -        & -       &&2&(2,3)&3&2&594.9 &$ -0.4\times 10^{-6}$ & $-401$ \\
  -        & -       &&2&(2,3)&5&3&685.0 &$ -0.4\times 10^{-4}$ & $-396$ \\
  -        & -       &&2&(2,3)&5&2&750.8 &$ -0.0003 $& $-392$ \\
 770.81(1) & 0.11(2) &&2&(2,3)&5&4&770.7&$ -0.5$ & $-390$ \\
 809.65(3) & 0.3(1)  &&2&(2,3)&3&3&809.7& $-0.09$& $-383$ \\
 819.8(2) & 1.1(5)  &&2&(2,3)&5&5&819.0& $-5.4$& $-380$ \\
 852.3(3) & 1.3(4)$^{\dagger}$  &&0&(2,3)&5&4&851.3 &$ -1199$ & $-393$ \\
  -        & -       &&2&(2,3)&2&2&961.8 & $-0.01$ & $-390$ \\
  -        & -       &&2&(2,3)&4&4&980.5 & $-0.7$&$ -387$ \\
\hline
\hline
\end{tabular}
\end{center}
\caption[]{Location and assignment of Feshbach resonances for $^{85}$Rb$_2$ in
the aa channel in the field range between 0 and 1000~G. All quantum numbers in
the table refer to the molecular states. The experimental errors shown are
statistical uncertainties resulting from the fits as described in the text. The
experimental width labeled $^{\dagger}$ was determined from the difference
between the minima and maxima in the measured atom number. Additional
systematic uncertainties of $0.1$~G and $0.5$~G apply to resonance positions in
the field ranges 0 to 400~G and 400 to 1000~G respectively.}
\label{table:85RbFeshbachResonances_aa}
\end{table}
\begin{table}[!]
\begin{center}
\begin{tabular}{ccccccccc}
\hline
\hline
\multicolumn{9}{c} {Incoming s-wave (2,$-2$)+(2,$-2$) state} \\
\hline
 \multicolumn{2}{c} {Experiment}& & \multicolumn{6}{c} {Theory}\\
 \cline{1-2}
 \cline{4-9}
 $B_0$ & $\delta$ & & \multicolumn{2}{c} {Assignment} & $B_0$ & $\Delta$ & $a_{\rm{res}}$ & $a_{\rm{bg}}$\\
 \cline{4-5}
  (G) & (G) &  & $L$ & $M_{F}$ & (G) & (mG) & (bohr) & (bohr) \\
\hline
 156(1) &  10.5(5) &&0&$-4$& 155.3& 10900 & $\ge$10000 & $-441$  \\
  -     & -        &&2&$-6$& 215.5 & 5.5 & 4000 & $-374$ \\
 219.58(1)& 0.22(9) &&0&$-4$& 219.9& 9.1 & 4000 & $-379$  \\
 232.25(1)& 0.23(1) &&2&$-4$& 232.5 & 2.0 & 400 & $-393$ \\
 248.64(1)& 0.12(2) &&2&$-5$& 248.9 & 2.9 & 5000 & $-406 $ \\
 297.42(1)& 0.09(1) &&2&$-4$& 297.7 & 1.8 & 5000 & $-432$ \\
 382.36(2)& 0.19(1) &&2 &$-3$& 382   &  -   &  15  & $-457 $\\
 532.3(3) & 3.2(1)$^{\dagger}$ &&0&$-4$&532.9& 2300 & $\ge$10000 & $-474$ \\
 604.1(1) & 0.2(1)  &&2&$-4$& 604.4  & 0.03 & 700 &$-466$  \\
  -       & -       &&2&$-5$& 854.3  & 0.002 & 25 & $-481$ \\
 924.52(4)& 2.8(1)  &&2&$-3$& 924    &  -   &  9
  & $-476$\\
\hline
\hline
\end{tabular}
\end{center}
\caption[]{Location and assignment of Feshbach resonances for $^{85}$Rb$_2$ in
the ee channel in the field range between 0 and 1000~G. All resonances shown
satisfy $a_{\rm{res}}\ge 1$~$a_0$. All quantum numbers in the table refer to
the molecular states. The experimental errors shown are statistical
uncertainties resulting from the fits as described in the text. The
experimental width labeled $^{\dagger}$ was determined from the difference
between the minima and maxima in the measured atom number. Additional
systematic uncertainties of 0.1~G and 0.5~G apply to resonance positions in the
field ranges 0 to 400~G and 400 to 1000~G respectively. The resonances near
155~G and 220~G have been measured previously \cite{Claussen:2003,Altin:2010b}.
} \label{table:85RbFeshbachResonances_ee}
\end{table}

The two inelastically dominated features that are seen experimentally are those
with the largest $a_{\rm res}$ values. The number of atoms in the trap
decreases around these resonances due to an increase in the 2-body loss rate,
as shown in Figure \ref{fig:Inelastic}(a) and (b). The inelastic collisions
also lead to an increase in temperature, as shown in Figure \ref{fig:Inelastic}
(c) and (d). The rate coefficient for 2-body losses due to inelastic collisions
from a channel $n$ is
\begin{equation}
K^{(2)}_{\rm loss}(B)=\frac{2h}{\mu}g_n\beta(B),
\end{equation}
where $\beta(B)$ is the imaginary part of the scattering length and $g_n=2$ for
a thermal cloud of identical bosons \cite{Chin:RMP:2010}. The calculated rate
coefficients for the two resonances are shown in panels (e) and (f); they peak
around $1\times10^{-11}$ cm$^{3}$/s, which is an order of magnitude higher than
for any of the other inelastically dominated features.

\begin{figure}[t]
\includegraphics[width=0.5\columnwidth]{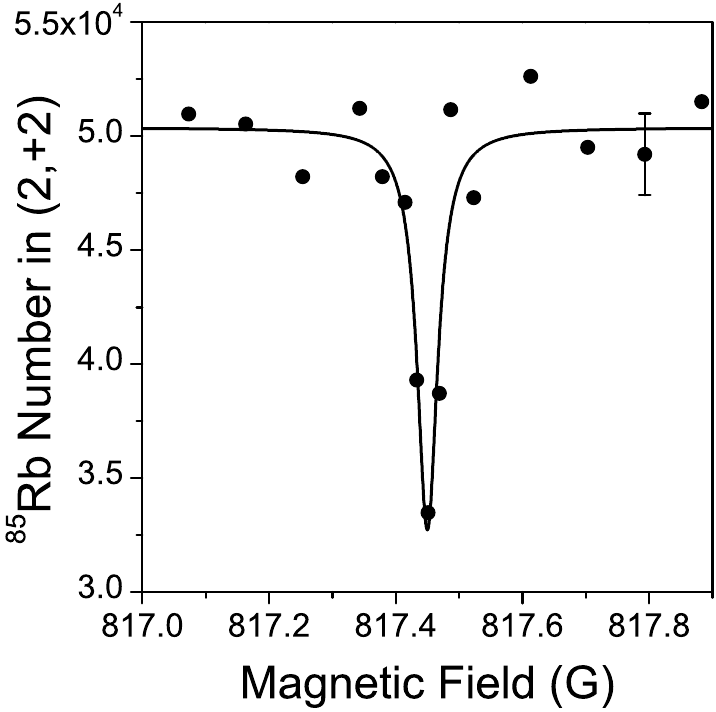}
\caption{A resonance measured between the $(2,+2)$ and $(3,+3)$
spin states in $^{85}$Rb at 817.45(5)~G. On resonance the increased inelastic
collision rate in the mixture results in a loss feature in the $(2,+2)$ atom
number as a function of magnetic field. The error bars show the standard
deviation for multiple control shots at a specific magnetic field.}
\label{fig:SpinMix}
\end{figure}

We have also measured one resonance in the $(2,+2)+(3,+3)$ mixed spin channel.
The experimental results are presented in Figure \ref{fig:SpinMix} where a loss
feature in the $^{85}$Rb $(2,+2)$ number reveals the location of the resonance.
A Lorentzian fit gives a resonance position of 817.45(5)~G and an experimental
width of 0.031(1)~G, which may be compared with the predicted position 818.8~G.

\begin{figure}[t]
\includegraphics[width=1\columnwidth]{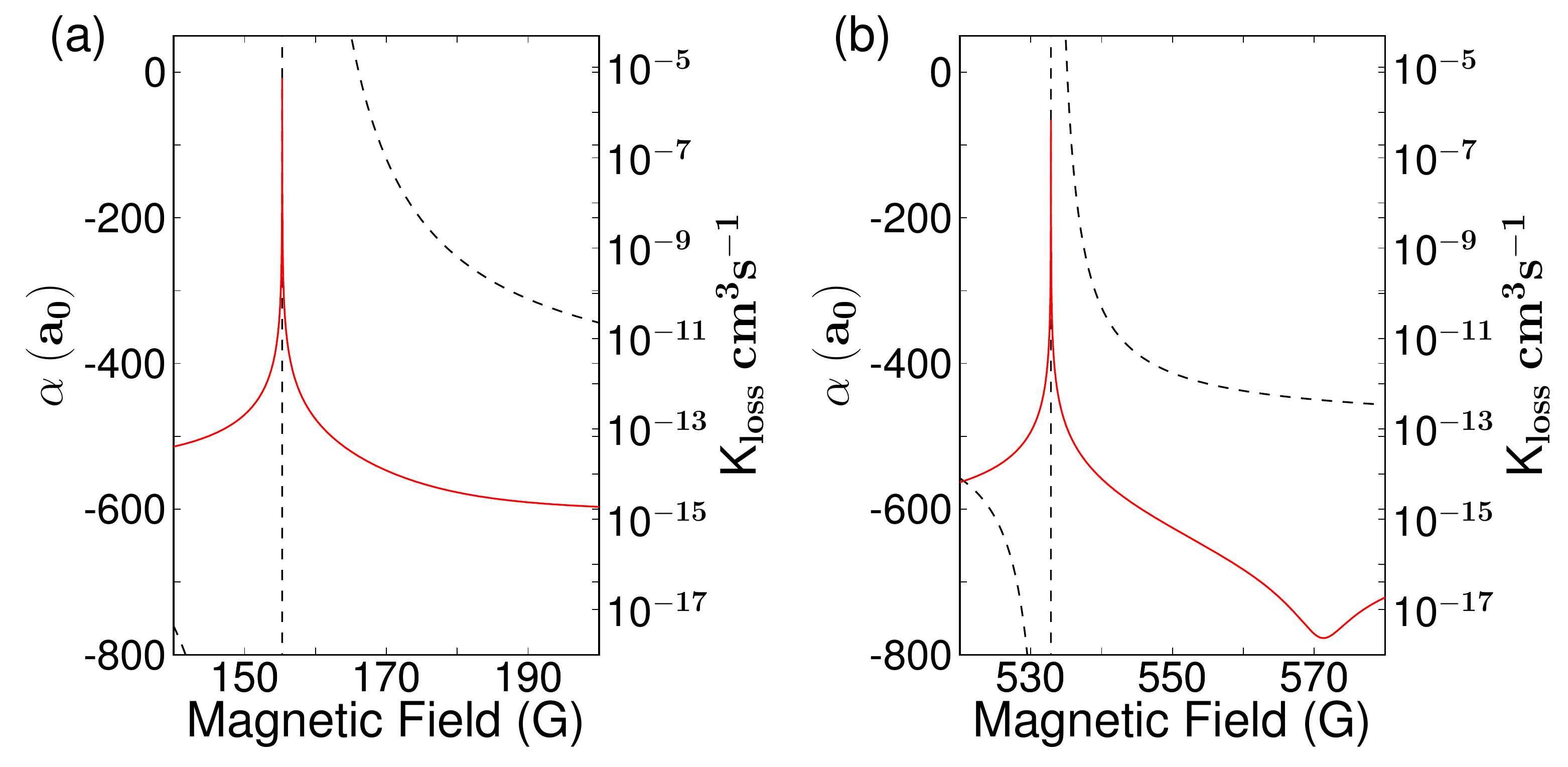}
\caption{The rate coefficient for 2-body loss $K_{\rm loss}$ (red solid lines),
which is proportional to the imaginary part of the scattering length, and the
corresponding real part of the scattering length (dashed lines) for the two
resonances with $\Delta > 1$~G in the ee channel. Note the dip in $K_{\rm
loss}$ on the high-field side the 532 G resonance.} \label{fig:kloss}
\end{figure}

\section{Conclusion}
A detailed understanding of the two-body scattering behavior is essential for
understanding many phenomena in ultracold gases. These include studies of
molecule formation \cite{Donley:2002, Thompson:magres:2005, Kohler:2005,
Hodby:2005, Brown:2006}, Efimov states and their universality \cite{Stoll:2005,
Altin:2011, Wild:2012, Berninger:Efimov:2011}, dimer collisions and few-body
physics \cite{Ferlaino:halo:2008}, BEC production \cite{Cornish:2000,
Weber:2003}, controlled condensate collapse \cite{Roberts:collapse:2001,
Donley:2001, Altin:2011}, and the formation of bright matter wave solitons
\cite{Cornish:2006}. The scattering properties of many alkali-metal atoms have
been documented in the literature \cite{Chin:RMP:2010}. However, nearly all
previous work on $^{85}$Rb has focussed on a single broad resonance near 155~G.
This paper redresses this balance by presenting a detailed study of the
scattering properties of $^{85}$Rb, revealing additional broad resonances and
numerous unreported narrow resonances in both the ee and aa channels. As has
been the case for other alkali-metal atoms, this work will facilitate many
future studies using $^{85}$Rb.

We have recently explored interspecies Feshbach resonances in mixtures of
$^{85}$Rb and $^{133}$Cs \cite{Cho:RbCs:2012}, as a key step towards the
production of $^{85}$Rb$^{133}$Cs molecules. The improved understanding of the
collisional behavior of $^{85}$Rb resulting from the present work is essential
for the production of the high phase-space density mixtures required for
efficient molecule formation. In particular, the two previously unmeasured
broad resonances presented here offer new magnetic field regions for
evaporative cooling. The elastic to inelastic collision ratio in the vicinity
of these features is potentially more favorable for evaporative cooling than
near the 155~G resonance, where direct evaporation of $^{85}$Rb to BEC is
possible \cite{Cornish:2000,Marchant:2012}. Figure \ref{fig:kloss} compares the
scattering properties around the 532~G resonance with those near the 155~G
resonance. The results for the 562~G resonance show a pronounced dip in the
rate coefficient for 2-body loss near 570~G, due to interference between the
resonant and background contributions to the inelastic scattering
\cite{Hutson:res:2007, Hutson:HeO2:2009}, which offers a range of magnetic
fields where more efficient cooling may be possible. No such dip in the 2-body
loss rate is present near the 155~G resonance. Alternatively, the aa channel
offers the prospect of evaporative cooling free from two-body loss. Although
the background scattering length is moderately large and negative for
ground-state atoms (see Figure \ref{fig:aa_scattering}), the broad resonance at
851~G may be used to tune the scattering length to modest positive values,
improving the evaporation efficiency and offering the prospect of BEC formation
directly in absolute ground state. In the future we will investigate
evaporative cooling in these new field regions. We will also use the improved
knowledge of the scattering of $^{85}$Rb presented in this paper, together with
similar knowledge for $^{133}$Cs \cite{Chin:cs2-fesh:2004,Berninger:Cs2:2013},
to devise a route to cooling $^{85}$Rb-$^{133}$Cs mixtures to suitable
phase-space densities for magneto-association using a narrow interspecies
resonance \cite{Cho:RbCs:2012}.

\section{Acknowledgements}
We thank Paul S. Julienne for many valuable discussions. This work was
supported by EPSRC and by EOARD Grant FA8655-10-1-3033. CLB is supported by a
Doctoral Fellowship from Durham University.

\bibliography{85Rb2_paper,../all}

%merlin.mbs apsrev4-1.bst 2010-07-25 4.21a (PWD, AO, DPC) hacked
%Control: key (0)
%Control: author (8) initials jnrlst
%Control: editor formatted (1) identically to author
%Control: production of article title (-1) disabled
%Control: page (0) single
%Control: year (1) truncated
%Control: production of eprint (0) enabled
\begin{thebibliography}{50}%
\makeatletter
\providecommand \@ifxundefined [1]{%
 \@ifx{#1\undefined}
}%
\providecommand \@ifnum [1]{%
 \ifnum #1\expandafter \@firstoftwo
 \else \expandafter \@secondoftwo
 \fi
}%
\providecommand \@ifx [1]{%
 \ifx #1\expandafter \@firstoftwo
 \else \expandafter \@secondoftwo
 \fi
}%
\providecommand \natexlab [1]{#1}%
\providecommand \enquote  [1]{``#1''}%
\providecommand \bibnamefont  [1]{#1}%
\providecommand \bibfnamefont [1]{#1}%
\providecommand \citenamefont [1]{#1}%
\providecommand \href@noop [0]{\@secondoftwo}%
\providecommand \href [0]{\begingroup \@sanitize@url \@href}%
\providecommand \@href[1]{\@@startlink{#1}\@@href}%
\providecommand \@@href[1]{\endgroup#1\@@endlink}%
\providecommand \@sanitize@url [0]{\catcode `\\12\catcode `\$12\catcode
  `\&12\catcode `\#12\catcode `\^12\catcode `\_12\catcode `\%12\relax}%
\providecommand \@@startlink[1]{}%
\providecommand \@@endlink[0]{}%
\providecommand \url  [0]{\begingroup\@sanitize@url \@url }%
\providecommand \@url [1]{\endgroup\@href {#1}{\urlprefix }}%
\providecommand \urlprefix  [0]{URL }%
\providecommand \Eprint [0]{\href }%
\providecommand \doibase [0]{http://dx.doi.org/}%
\providecommand \selectlanguage [0]{\@gobble}%
\providecommand \bibinfo  [0]{\@secondoftwo}%
\providecommand \bibfield  [0]{\@secondoftwo}%
\providecommand \translation [1]{[#1]}%
\providecommand \BibitemOpen [0]{}%
\providecommand \bibitemStop [0]{}%
\providecommand \bibitemNoStop [0]{.\EOS\space}%
\providecommand \EOS [0]{\spacefactor3000\relax}%
\providecommand \BibitemShut  [1]{\csname bibitem#1\endcsname}%
\let\auto@bib@innerbib\@empty
%</preamble>
\bibitem [{\citenamefont {Carr}\ \emph {et~al.}(2009)\citenamefont {Carr},
  \citenamefont {{DeMille}}, \citenamefont {Krems},\ and\ \citenamefont
  {Ye}}]{Carr:NJPintro:2009}%
  \BibitemOpen
  \bibfield  {author} {\bibinfo {author} {\bibfnamefont {L.~D.}\ \bibnamefont
  {Carr}}, \bibinfo {author} {\bibfnamefont {D.}~\bibnamefont {{DeMille}}},
  \bibinfo {author} {\bibfnamefont {R.~V.}\ \bibnamefont {Krems}}, \ and\
  \bibinfo {author} {\bibfnamefont {J.}~\bibnamefont {Ye}},\ }\href@noop {}
  {\bibfield  {journal} {\bibinfo  {journal} {New J. Phys.}\ }\textbf {\bibinfo
  {volume} {11}},\ \bibinfo {pages} {055049} (\bibinfo {year}
  {2009})}\BibitemShut {NoStop}%
\bibitem [{\citenamefont {Krems}(2005)}]{Krems:IRPC:2005}%
  \BibitemOpen
  \bibfield  {author} {\bibinfo {author} {\bibfnamefont {R.~V.}\ \bibnamefont
  {Krems}},\ }\href@noop {} {\bibfield  {journal} {\bibinfo  {journal} {Int.
  Rev. Phys. Chem.}\ }\textbf {\bibinfo {volume} {24}},\ \bibinfo {pages} {99}
  (\bibinfo {year} {2005})}\BibitemShut {NoStop}%
\bibitem [{\citenamefont {Chin}\ \emph {et~al.}(2010)\citenamefont {Chin},
  \citenamefont {Grimm}, \citenamefont {Tiesinga},\ and\ \citenamefont
  {Julienne}}]{Chin:RMP:2010}%
  \BibitemOpen
  \bibfield  {author} {\bibinfo {author} {\bibfnamefont {C.}~\bibnamefont
  {Chin}}, \bibinfo {author} {\bibfnamefont {R.}~\bibnamefont {Grimm}},
  \bibinfo {author} {\bibfnamefont {E.}~\bibnamefont {Tiesinga}}, \ and\
  \bibinfo {author} {\bibfnamefont {P.~S.}\ \bibnamefont {Julienne}},\
  }\href@noop {} {\bibfield  {journal} {\bibinfo  {journal} {Rev. Mod. Phys.}\
  }\textbf {\bibinfo {volume} {82}},\ \bibinfo {pages} {1225} (\bibinfo {year}
  {2010})}\BibitemShut {NoStop}%
\bibitem [{\citenamefont {Damski}\ \emph {et~al.}(2003)\citenamefont {Damski},
  \citenamefont {Santos}, \citenamefont {Tiemann}, \citenamefont {Lewenstein},
  \citenamefont {Kotochigova}, \citenamefont {Julienne},\ and\ \citenamefont
  {Zoller}}]{Damski:2003}%
  \BibitemOpen
  \bibfield  {author} {\bibinfo {author} {\bibfnamefont {B.}~\bibnamefont
  {Damski}}, \bibinfo {author} {\bibfnamefont {L.}~\bibnamefont {Santos}},
  \bibinfo {author} {\bibfnamefont {E.}~\bibnamefont {Tiemann}}, \bibinfo
  {author} {\bibfnamefont {M.}~\bibnamefont {Lewenstein}}, \bibinfo {author}
  {\bibfnamefont {S.}~\bibnamefont {Kotochigova}}, \bibinfo {author}
  {\bibfnamefont {P.}~\bibnamefont {Julienne}}, \ and\ \bibinfo {author}
  {\bibfnamefont {P.}~\bibnamefont {Zoller}},\ }\href@noop {} {\bibfield
  {journal} {\bibinfo  {journal} {Phys. Rev. Lett.}\ }\textbf {\bibinfo
  {volume} {90}},\ \bibinfo {pages} {110401} (\bibinfo {year}
  {2003})}\BibitemShut {NoStop}%
\bibitem [{\citenamefont {Ni}\ \emph {et~al.}(2008)\citenamefont {Ni},
  \citenamefont {Ospelkaus}, \citenamefont {{de Miranda}}, \citenamefont
  {Pe'er}, \citenamefont {Neyenhuis}, \citenamefont {Zirbel}, \citenamefont
  {Kotochigova}, \citenamefont {Julienne}, \citenamefont {Jin},\ and\
  \citenamefont {Ye}}]{Ni:KRb:2008}%
  \BibitemOpen
  \bibfield  {author} {\bibinfo {author} {\bibfnamefont {K.-K.}\ \bibnamefont
  {Ni}}, \bibinfo {author} {\bibfnamefont {S.}~\bibnamefont {Ospelkaus}},
  \bibinfo {author} {\bibfnamefont {M.~H.~G.}\ \bibnamefont {{de Miranda}}},
  \bibinfo {author} {\bibfnamefont {A.}~\bibnamefont {Pe'er}}, \bibinfo
  {author} {\bibfnamefont {B.}~\bibnamefont {Neyenhuis}}, \bibinfo {author}
  {\bibfnamefont {J.~J.}\ \bibnamefont {Zirbel}}, \bibinfo {author}
  {\bibfnamefont {S.}~\bibnamefont {Kotochigova}}, \bibinfo {author}
  {\bibfnamefont {P.~S.}\ \bibnamefont {Julienne}}, \bibinfo {author}
  {\bibfnamefont {D.~S.}\ \bibnamefont {Jin}}, \ and\ \bibinfo {author}
  {\bibfnamefont {J.}~\bibnamefont {Ye}},\ }\href@noop {} {\bibfield  {journal}
  {\bibinfo  {journal} {Science}\ }\textbf {\bibinfo {volume} {322}},\ \bibinfo
  {pages} {231} (\bibinfo {year} {2008})}\BibitemShut {NoStop}%
\bibitem [{\citenamefont {Lang}\ \emph {et~al.}(2008)\citenamefont {Lang},
  \citenamefont {Winkler}, \citenamefont {Strauss}, \citenamefont {Grimm},\
  and\ \citenamefont {Hecker~Denschlag}}]{Lang:ground:2008}%
  \BibitemOpen
  \bibfield  {author} {\bibinfo {author} {\bibfnamefont {F.}~\bibnamefont
  {Lang}}, \bibinfo {author} {\bibfnamefont {K.}~\bibnamefont {Winkler}},
  \bibinfo {author} {\bibfnamefont {C.}~\bibnamefont {Strauss}}, \bibinfo
  {author} {\bibfnamefont {R.}~\bibnamefont {Grimm}}, \ and\ \bibinfo {author}
  {\bibfnamefont {J.}~\bibnamefont {Hecker~Denschlag}},\ }\href@noop {}
  {\bibfield  {journal} {\bibinfo  {journal} {Phys. Rev. Lett.}\ }\textbf
  {\bibinfo {volume} {101}},\ \bibinfo {pages} {133005} (\bibinfo {year}
  {2008})}\BibitemShut {NoStop}%
\bibitem [{\citenamefont {Danzl}\ \emph {et~al.}(2010)\citenamefont {Danzl},
  \citenamefont {Mark}, \citenamefont {Haller}, \citenamefont {Gustavsson},
  \citenamefont {Hart}, \citenamefont {Aldegunde}, \citenamefont {Hutson},\
  and\ \citenamefont {N\"agerl}}]{Danzl:ground:2010}%
  \BibitemOpen
  \bibfield  {author} {\bibinfo {author} {\bibfnamefont {J.~G.}\ \bibnamefont
  {Danzl}}, \bibinfo {author} {\bibfnamefont {M.~J.}\ \bibnamefont {Mark}},
  \bibinfo {author} {\bibfnamefont {E.}~\bibnamefont {Haller}}, \bibinfo
  {author} {\bibfnamefont {M.}~\bibnamefont {Gustavsson}}, \bibinfo {author}
  {\bibfnamefont {R.}~\bibnamefont {Hart}}, \bibinfo {author} {\bibfnamefont
  {J.}~\bibnamefont {Aldegunde}}, \bibinfo {author} {\bibfnamefont {J.~M.}\
  \bibnamefont {Hutson}}, \ and\ \bibinfo {author} {\bibfnamefont {H.-C.}\
  \bibnamefont {N\"agerl}},\ }\href {\doibase doi:10.1038/nphys1533} {\bibfield
   {journal} {\bibinfo  {journal} {Nature Phys.}\ }\textbf {\bibinfo {volume}
  {6}},\ \bibinfo {pages} {265} (\bibinfo {year} {2010})}\BibitemShut {NoStop}%
\bibitem [{\citenamefont {Burke}\ \emph {et~al.}(1998)\citenamefont {Burke},
  \citenamefont {Bohn}, \citenamefont {Esry},\ and\ \citenamefont
  {Greene}}]{Burke:1998}%
  \BibitemOpen
  \bibfield  {author} {\bibinfo {author} {\bibfnamefont {J.~P.}\ \bibnamefont
  {Burke}}, \bibinfo {author} {\bibfnamefont {J.~L.}\ \bibnamefont {Bohn}},
  \bibinfo {author} {\bibfnamefont {B.~D.}\ \bibnamefont {Esry}}, \ and\
  \bibinfo {author} {\bibfnamefont {C.~H.}\ \bibnamefont {Greene}},\
  }\href@noop {} {\bibfield  {journal} {\bibinfo  {journal} {Phys. Rev. Lett.}\
  }\textbf {\bibinfo {volume} {80}},\ \bibinfo {pages} {2097} (\bibinfo {year}
  {1998})}\BibitemShut {NoStop}%
\bibitem [{\citenamefont {Cornish}\ \emph {et~al.}(2000)\citenamefont
  {Cornish}, \citenamefont {Claussen}, \citenamefont {Roberts}, \citenamefont
  {Cornell},\ and\ \citenamefont {Wieman}}]{Cornish:2000}%
  \BibitemOpen
  \bibfield  {author} {\bibinfo {author} {\bibfnamefont {S.~L.}\ \bibnamefont
  {Cornish}}, \bibinfo {author} {\bibfnamefont {N.~R.}\ \bibnamefont
  {Claussen}}, \bibinfo {author} {\bibfnamefont {J.~L.}\ \bibnamefont
  {Roberts}}, \bibinfo {author} {\bibfnamefont {E.~A.}\ \bibnamefont
  {Cornell}}, \ and\ \bibinfo {author} {\bibfnamefont {C.~E.}\ \bibnamefont
  {Wieman}},\ }\href@noop {} {\bibfield  {journal} {\bibinfo  {journal} {Phys.
  Rev. Lett.}\ }\textbf {\bibinfo {volume} {85}},\ \bibinfo {pages} {1795}
  (\bibinfo {year} {2000})}\BibitemShut {NoStop}%
\bibitem [{\citenamefont {Cho}\ \emph {et~al.}(2012)\citenamefont {Cho},
  \citenamefont {McCarron}, \citenamefont {K\"oppinger}, \citenamefont
  {Jenkin}, \citenamefont {Butler}, \citenamefont {Julienne}, \citenamefont
  {Blackley}, \citenamefont {{Le Sueur}}, \citenamefont {Hutson},\ and\
  \citenamefont {Cornish}}]{Cho:RbCs:2012}%
  \BibitemOpen
  \bibfield  {author} {\bibinfo {author} {\bibfnamefont {H.-W.}\ \bibnamefont
  {Cho}}, \bibinfo {author} {\bibfnamefont {D.~J.}\ \bibnamefont {McCarron}},
  \bibinfo {author} {\bibfnamefont {M.~P.}\ \bibnamefont {K\"oppinger}},
  \bibinfo {author} {\bibfnamefont {D.~L.}\ \bibnamefont {Jenkin}}, \bibinfo
  {author} {\bibfnamefont {K.~L.}\ \bibnamefont {Butler}}, \bibinfo {author}
  {\bibfnamefont {P.~S.}\ \bibnamefont {Julienne}}, \bibinfo {author}
  {\bibfnamefont {C.~L.}\ \bibnamefont {Blackley}}, \bibinfo {author}
  {\bibfnamefont {C.~R.}\ \bibnamefont {{Le Sueur}}}, \bibinfo {author}
  {\bibfnamefont {J.~M.}\ \bibnamefont {Hutson}}, \ and\ \bibinfo {author}
  {\bibfnamefont {S.~L.}\ \bibnamefont {Cornish}},\ }\href@noop {} {\bibfield
  {journal} {\bibinfo  {journal} {arXiv:1208.4569}\ } (\bibinfo {year}
  {2012})}\BibitemShut {NoStop}%
\bibitem [{\citenamefont {Vogels}\ \emph {et~al.}(1997)\citenamefont {Vogels},
  \citenamefont {Tsai}, \citenamefont {Freeland}, \citenamefont {Kokkelmans},
  \citenamefont {Verhaar},\ and\ \citenamefont {Heinzen}}]{Vogels:1997}%
  \BibitemOpen
  \bibfield  {author} {\bibinfo {author} {\bibfnamefont {J.~M.}\ \bibnamefont
  {Vogels}}, \bibinfo {author} {\bibfnamefont {C.~C.}\ \bibnamefont {Tsai}},
  \bibinfo {author} {\bibfnamefont {R.~S.}\ \bibnamefont {Freeland}}, \bibinfo
  {author} {\bibfnamefont {S.~J. J. M.~F.}\ \bibnamefont {Kokkelmans}},
  \bibinfo {author} {\bibfnamefont {B.~J.}\ \bibnamefont {Verhaar}}, \ and\
  \bibinfo {author} {\bibfnamefont {D.~J.}\ \bibnamefont {Heinzen}},\ }\href
  {\doibase 10.1103/PhysRevA.56.R1067} {\bibfield  {journal} {\bibinfo
  {journal} {Phys. Rev. A}\ }\textbf {\bibinfo {volume} {56}},\ \bibinfo
  {pages} {R1067} (\bibinfo {year} {1997})}\BibitemShut {NoStop}%
\bibitem [{\citenamefont {Roberts}\ \emph
  {et~al.}(2001{\natexlab{a}})\citenamefont {Roberts}, \citenamefont
  {Claussen}, \citenamefont {Cornish}, \citenamefont {Donley}, \citenamefont
  {Cornell},\ and\ \citenamefont {Wieman}}]{Roberts:collapse:2001}%
  \BibitemOpen
  \bibfield  {author} {\bibinfo {author} {\bibfnamefont {J.~L.}\ \bibnamefont
  {Roberts}}, \bibinfo {author} {\bibfnamefont {N.~R.}\ \bibnamefont
  {Claussen}}, \bibinfo {author} {\bibfnamefont {S.~L.}\ \bibnamefont
  {Cornish}}, \bibinfo {author} {\bibfnamefont {E.~A.}\ \bibnamefont {Donley}},
  \bibinfo {author} {\bibfnamefont {E.~A.}\ \bibnamefont {Cornell}}, \ and\
  \bibinfo {author} {\bibfnamefont {C.~E.}\ \bibnamefont {Wieman}},\
  }\href@noop {} {\bibfield  {journal} {\bibinfo  {journal} {Phys. Rev. Lett.}\
  }\textbf {\bibinfo {volume} {86}},\ \bibinfo {pages} {4211} (\bibinfo {year}
  {2001}{\natexlab{a}})}\BibitemShut {NoStop}%
\bibitem [{\citenamefont {Donley}\ \emph {et~al.}(2001)\citenamefont {Donley},
  \citenamefont {Claussen}, \citenamefont {Cornish}, \citenamefont {Roberts},
  \citenamefont {Cornell},\ and\ \citenamefont {Wieman}}]{Donley:2001}%
  \BibitemOpen
  \bibfield  {author} {\bibinfo {author} {\bibfnamefont {E.~A.}\ \bibnamefont
  {Donley}}, \bibinfo {author} {\bibfnamefont {N.~R.}\ \bibnamefont
  {Claussen}}, \bibinfo {author} {\bibfnamefont {S.~L.}\ \bibnamefont
  {Cornish}}, \bibinfo {author} {\bibfnamefont {J.~L.}\ \bibnamefont
  {Roberts}}, \bibinfo {author} {\bibfnamefont {E.~A.}\ \bibnamefont
  {Cornell}}, \ and\ \bibinfo {author} {\bibfnamefont {C.~E.}\ \bibnamefont
  {Wieman}},\ }\href@noop {} {\bibfield  {journal} {\bibinfo  {journal}
  {Nature}\ }\textbf {\bibinfo {volume} {412}},\ \bibinfo {pages} {295}
  (\bibinfo {year} {2001})}\BibitemShut {NoStop}%
\bibitem [{\citenamefont {Altin}\ \emph {et~al.}(2011)\citenamefont {Altin},
  \citenamefont {Dennis}, \citenamefont {McDonald}, \citenamefont {D\"oring},
  \citenamefont {Debs}, \citenamefont {Close}, \citenamefont {Savage},\ and\
  \citenamefont {Robins}}]{Altin:2011}%
  \BibitemOpen
  \bibfield  {author} {\bibinfo {author} {\bibfnamefont {P.~A.}\ \bibnamefont
  {Altin}}, \bibinfo {author} {\bibfnamefont {G.~R.}\ \bibnamefont {Dennis}},
  \bibinfo {author} {\bibfnamefont {G.~D.}\ \bibnamefont {McDonald}}, \bibinfo
  {author} {\bibfnamefont {D.}~\bibnamefont {D\"oring}}, \bibinfo {author}
  {\bibfnamefont {J.~E.}\ \bibnamefont {Debs}}, \bibinfo {author}
  {\bibfnamefont {J.~D.}\ \bibnamefont {Close}}, \bibinfo {author}
  {\bibfnamefont {C.~M.}\ \bibnamefont {Savage}}, \ and\ \bibinfo {author}
  {\bibfnamefont {N.~P.}\ \bibnamefont {Robins}},\ }\href {\doibase
  10.1103/PhysRevA.84.033632} {\bibfield  {journal} {\bibinfo  {journal} {Phys.
  Rev. A}\ }\textbf {\bibinfo {volume} {84}},\ \bibinfo {pages} {033632}
  (\bibinfo {year} {2011})}\BibitemShut {NoStop}%
\bibitem [{\citenamefont {Cornish}\ \emph {et~al.}(2006)\citenamefont
  {Cornish}, \citenamefont {Thompson},\ and\ \citenamefont
  {Wieman}}]{Cornish:2006}%
  \BibitemOpen
  \bibfield  {author} {\bibinfo {author} {\bibfnamefont {S.~L.}\ \bibnamefont
  {Cornish}}, \bibinfo {author} {\bibfnamefont {S.~T.}\ \bibnamefont
  {Thompson}}, \ and\ \bibinfo {author} {\bibfnamefont {C.~E.}\ \bibnamefont
  {Wieman}},\ }\href {\doibase 10.1103/PhysRevLett.96.170401} {\bibfield
  {journal} {\bibinfo  {journal} {Phys. Rev. Lett.}\ }\textbf {\bibinfo
  {volume} {96}},\ \bibinfo {pages} {170401} (\bibinfo {year}
  {2006})}\BibitemShut {NoStop}%
\bibitem [{\citenamefont {Wild}\ \emph {et~al.}(2012)\citenamefont {Wild},
  \citenamefont {Makotyn}, \citenamefont {Pino}, \citenamefont {Cornell},\ and\
  \citenamefont {Jin}}]{Wild:2012}%
  \BibitemOpen
  \bibfield  {author} {\bibinfo {author} {\bibfnamefont {R.~J.}\ \bibnamefont
  {Wild}}, \bibinfo {author} {\bibfnamefont {P.}~\bibnamefont {Makotyn}},
  \bibinfo {author} {\bibfnamefont {J.~M.}\ \bibnamefont {Pino}}, \bibinfo
  {author} {\bibfnamefont {E.~A.}\ \bibnamefont {Cornell}}, \ and\ \bibinfo
  {author} {\bibfnamefont {D.~S.}\ \bibnamefont {Jin}},\ }\href {\doibase
  10.1103/PhysRevLett.108.145305} {\bibfield  {journal} {\bibinfo  {journal}
  {Phys. Rev. Lett.}\ }\textbf {\bibinfo {volume} {108}},\ \bibinfo {pages}
  {145305} (\bibinfo {year} {2012})}\BibitemShut {NoStop}%
\bibitem [{\citenamefont {Tsai}\ \emph {et~al.}(1997)\citenamefont {Tsai},
  \citenamefont {Freeland}, \citenamefont {Vogels}, \citenamefont {Boesten},
  \citenamefont {Verhaar},\ and\ \citenamefont {Heinzen}}]{Tsai:1997}%
  \BibitemOpen
  \bibfield  {author} {\bibinfo {author} {\bibfnamefont {C.~C.}\ \bibnamefont
  {Tsai}}, \bibinfo {author} {\bibfnamefont {R.~S.}\ \bibnamefont {Freeland}},
  \bibinfo {author} {\bibfnamefont {J.~M.}\ \bibnamefont {Vogels}}, \bibinfo
  {author} {\bibfnamefont {H.~M. J.~M.}\ \bibnamefont {Boesten}}, \bibinfo
  {author} {\bibfnamefont {B.~J.}\ \bibnamefont {Verhaar}}, \ and\ \bibinfo
  {author} {\bibfnamefont {D.~J.}\ \bibnamefont {Heinzen}},\ }\href {\doibase
  10.1103/PhysRevLett.79.1245} {\bibfield  {journal} {\bibinfo  {journal}
  {Phys. Rev. Lett.}\ }\textbf {\bibinfo {volume} {79}},\ \bibinfo {pages}
  {1245} (\bibinfo {year} {1997})}\BibitemShut {NoStop}%
\bibitem [{\citenamefont {Courteille}\ \emph {et~al.}(1998)\citenamefont
  {Courteille}, \citenamefont {Freeland}, \citenamefont {Heinzen},
  \citenamefont {{van Abeelen}},\ and\ \citenamefont
  {Verhaar}}]{Courteille:1998}%
  \BibitemOpen
  \bibfield  {author} {\bibinfo {author} {\bibfnamefont {P.}~\bibnamefont
  {Courteille}}, \bibinfo {author} {\bibfnamefont {R.~S.}\ \bibnamefont
  {Freeland}}, \bibinfo {author} {\bibfnamefont {D.~J.}\ \bibnamefont
  {Heinzen}}, \bibinfo {author} {\bibfnamefont {F.~A.}\ \bibnamefont {{van
  Abeelen}}}, \ and\ \bibinfo {author} {\bibfnamefont {B.~J.}\ \bibnamefont
  {Verhaar}},\ }\href@noop {} {\bibfield  {journal} {\bibinfo  {journal} {Phys.
  Rev. Lett.}\ }\textbf {\bibinfo {volume} {81}},\ \bibinfo {pages} {69}
  (\bibinfo {year} {1998})}\BibitemShut {NoStop}%
\bibitem [{\citenamefont {Roberts}\ \emph {et~al.}(2000)\citenamefont
  {Roberts}, \citenamefont {Claussen}, \citenamefont {Cornish},\ and\
  \citenamefont {Wieman}}]{Roberts:2000}%
  \BibitemOpen
  \bibfield  {author} {\bibinfo {author} {\bibfnamefont {J.~L.}\ \bibnamefont
  {Roberts}}, \bibinfo {author} {\bibfnamefont {N.~R.}\ \bibnamefont
  {Claussen}}, \bibinfo {author} {\bibfnamefont {S.~L.}\ \bibnamefont
  {Cornish}}, \ and\ \bibinfo {author} {\bibfnamefont {C.~E.}\ \bibnamefont
  {Wieman}},\ }\href@noop {} {\bibfield  {journal} {\bibinfo  {journal} {Phys.
  Rev. Lett.}\ }\textbf {\bibinfo {volume} {85}},\ \bibinfo {pages} {728}
  (\bibinfo {year} {2000})}\BibitemShut {NoStop}%
\bibitem [{\citenamefont {Roberts}\ \emph
  {et~al.}(2001{\natexlab{b}})\citenamefont {Roberts}, \citenamefont {Burke},
  \citenamefont {Claussen}, \citenamefont {Cornish}, \citenamefont {Donley},\
  and\ \citenamefont {Wieman}}]{Roberts:elast:2001}%
  \BibitemOpen
  \bibfield  {author} {\bibinfo {author} {\bibfnamefont {J.~L.}\ \bibnamefont
  {Roberts}}, \bibinfo {author} {\bibfnamefont {J.~P.}\ \bibnamefont {Burke}},
  \bibinfo {author} {\bibfnamefont {N.~R.}\ \bibnamefont {Claussen}}, \bibinfo
  {author} {\bibfnamefont {S.~L.}\ \bibnamefont {Cornish}}, \bibinfo {author}
  {\bibfnamefont {E.~A.}\ \bibnamefont {Donley}}, \ and\ \bibinfo {author}
  {\bibfnamefont {C.~E.}\ \bibnamefont {Wieman}},\ }\href@noop {} {\bibfield
  {journal} {\bibinfo  {journal} {Phys. Rev. A}\ }\textbf {\bibinfo {volume}
  {64}},\ \bibinfo {pages} {024702} (\bibinfo {year}
  {2001}{\natexlab{b}})}\BibitemShut {NoStop}%
\bibitem [{\citenamefont {Claussen}\ \emph {et~al.}(2003)\citenamefont
  {Claussen}, \citenamefont {Kokkelmans}, \citenamefont {Thompson},
  \citenamefont {Donley}, \citenamefont {Hodby},\ and\ \citenamefont
  {Wieman}}]{Claussen:2003}%
  \BibitemOpen
  \bibfield  {author} {\bibinfo {author} {\bibfnamefont {N.~R.}\ \bibnamefont
  {Claussen}}, \bibinfo {author} {\bibfnamefont {S.~J. J. M.~F.}\ \bibnamefont
  {Kokkelmans}}, \bibinfo {author} {\bibfnamefont {S.~T.}\ \bibnamefont
  {Thompson}}, \bibinfo {author} {\bibfnamefont {E.~A.}\ \bibnamefont
  {Donley}}, \bibinfo {author} {\bibfnamefont {E.}~\bibnamefont {Hodby}}, \
  and\ \bibinfo {author} {\bibfnamefont {C.~E.}\ \bibnamefont {Wieman}},\
  }\href@noop {} {\bibfield  {journal} {\bibinfo  {journal} {Phys. Rev. A}\
  }\textbf {\bibinfo {volume} {67}},\ \bibinfo {pages} {060701} (\bibinfo
  {year} {2003})}\BibitemShut {NoStop}%
\bibitem [{\citenamefont {Donley}\ \emph {et~al.}(2002)\citenamefont {Donley},
  \citenamefont {Claussen}, \citenamefont {Thompson},\ and\ \citenamefont
  {Wieman}}]{Donley:2002}%
  \BibitemOpen
  \bibfield  {author} {\bibinfo {author} {\bibfnamefont {E.~A.}\ \bibnamefont
  {Donley}}, \bibinfo {author} {\bibfnamefont {N.~R.}\ \bibnamefont
  {Claussen}}, \bibinfo {author} {\bibfnamefont {S.~T.}\ \bibnamefont
  {Thompson}}, \ and\ \bibinfo {author} {\bibfnamefont {C.~E.}\ \bibnamefont
  {Wieman}},\ }\href@noop {} {\bibfield  {journal} {\bibinfo  {journal}
  {Nature}\ }\textbf {\bibinfo {volume} {417}},\ \bibinfo {pages} {529}
  (\bibinfo {year} {2002})}\BibitemShut {NoStop}%
\bibitem [{\citenamefont {Thompson}\ \emph {et~al.}(2005)\citenamefont
  {Thompson}, \citenamefont {Hodby},\ and\ \citenamefont
  {Wieman}}]{Thompson:magres:2005}%
  \BibitemOpen
  \bibfield  {author} {\bibinfo {author} {\bibfnamefont {S.~T.}\ \bibnamefont
  {Thompson}}, \bibinfo {author} {\bibfnamefont {E.}~\bibnamefont {Hodby}}, \
  and\ \bibinfo {author} {\bibfnamefont {C.~E.}\ \bibnamefont {Wieman}},\
  }\href@noop {} {\bibfield  {journal} {\bibinfo  {journal} {Phys. Rev. Lett.}\
  }\textbf {\bibinfo {volume} {95}},\ \bibinfo {pages} {190404} (\bibinfo
  {year} {2005})}\BibitemShut {NoStop}%
\bibitem [{\citenamefont {K\"ohler}\ \emph {et~al.}(2005)\citenamefont
  {K\"ohler}, \citenamefont {Tiesinga},\ and\ \citenamefont
  {Julienne}}]{Kohler:2005}%
  \BibitemOpen
  \bibfield  {author} {\bibinfo {author} {\bibfnamefont {T.}~\bibnamefont
  {K\"ohler}}, \bibinfo {author} {\bibfnamefont {E.}~\bibnamefont {Tiesinga}},
  \ and\ \bibinfo {author} {\bibfnamefont {P.~S.}\ \bibnamefont {Julienne}},\
  }\href@noop {} {\bibfield  {journal} {\bibinfo  {journal} {Phys. Rev. Lett.}\
  }\textbf {\bibinfo {volume} {94}},\ \bibinfo {pages} {020402} (\bibinfo
  {year} {2005})}\BibitemShut {NoStop}%
\bibitem [{\citenamefont {Hodby}\ \emph {et~al.}(2005)\citenamefont {Hodby},
  \citenamefont {Thompson}, \citenamefont {Regal}, \citenamefont {Greiner},
  \citenamefont {Wilson}, \citenamefont {Jin}, \citenamefont {Cornell},\ and\
  \citenamefont {Wieman}}]{Hodby:2005}%
  \BibitemOpen
  \bibfield  {author} {\bibinfo {author} {\bibfnamefont {E.}~\bibnamefont
  {Hodby}}, \bibinfo {author} {\bibfnamefont {S.~T.}\ \bibnamefont {Thompson}},
  \bibinfo {author} {\bibfnamefont {C.~A.}\ \bibnamefont {Regal}}, \bibinfo
  {author} {\bibfnamefont {M.}~\bibnamefont {Greiner}}, \bibinfo {author}
  {\bibfnamefont {A.~C.}\ \bibnamefont {Wilson}}, \bibinfo {author}
  {\bibfnamefont {D.~S.}\ \bibnamefont {Jin}}, \bibinfo {author} {\bibfnamefont
  {E.~A.}\ \bibnamefont {Cornell}}, \ and\ \bibinfo {author} {\bibfnamefont
  {C.~E.}\ \bibnamefont {Wieman}},\ }\href@noop {} {\bibfield  {journal}
  {\bibinfo  {journal} {Phys. Rev. Lett.}\ }\textbf {\bibinfo {volume} {94}},\
  \bibinfo {pages} {120402} (\bibinfo {year} {2005})}\BibitemShut {NoStop}%
\bibitem [{\citenamefont {Brown}\ \emph {et~al.}(2006)\citenamefont {Brown},
  \citenamefont {Dicks},\ and\ \citenamefont {Walmsley}}]{Brown:2006}%
  \BibitemOpen
  \bibfield  {author} {\bibinfo {author} {\bibfnamefont {B.~L.}\ \bibnamefont
  {Brown}}, \bibinfo {author} {\bibfnamefont {A.~J.}\ \bibnamefont {Dicks}}, \
  and\ \bibinfo {author} {\bibfnamefont {I.~A.}\ \bibnamefont {Walmsley}},\
  }\href@noop {} {\bibfield  {journal} {\bibinfo  {journal} {Phys. Rev. Lett.}\
  }\textbf {\bibinfo {volume} {96}},\ \bibinfo {pages} {173002} (\bibinfo
  {year} {2006})}\BibitemShut {NoStop}%
\bibitem [{\citenamefont {Stoll}\ and\ \citenamefont
  {K\"{o}hler}(2005)}]{Stoll:2005}%
  \BibitemOpen
  \bibfield  {author} {\bibinfo {author} {\bibfnamefont {M.}~\bibnamefont
  {Stoll}}\ and\ \bibinfo {author} {\bibfnamefont {T.}~\bibnamefont
  {K\"{o}hler}},\ }\href@noop {} {\bibfield  {journal} {\bibinfo  {journal}
  {Phys. Rev. A}\ }\textbf {\bibinfo {volume} {72}},\ \bibinfo {pages} {022714}
  (\bibinfo {year} {2005})}\BibitemShut {NoStop}%
\bibitem [{\citenamefont {Hutson}\ and\ \citenamefont
  {Green}(1994)}]{molscat:v14}%
  \BibitemOpen
  \bibfield  {author} {\bibinfo {author} {\bibfnamefont {J.~M.}\ \bibnamefont
  {Hutson}}\ and\ \bibinfo {author} {\bibfnamefont {S.}~\bibnamefont {Green}},\
  }\href@noop {} {\enquote {\bibinfo {title} {{MOLSCAT} computer program,
  version 14},}\ }\bibinfo {howpublished} {distributed by Collaborative
  Computational Project No.\ 6 of the UK Engineering and Physical Sciences
  Research Council} (\bibinfo {year} {1994})\BibitemShut {NoStop}%
\bibitem [{\citenamefont {Gonz\'{a}lez-Mart\'{\i}nez}\ and\ \citenamefont
  {Hutson}(2007)}]{Gonzalez-Martinez:2007}%
  \BibitemOpen
  \bibfield  {author} {\bibinfo {author} {\bibfnamefont {M.~L.}\ \bibnamefont
  {Gonz\'{a}lez-Mart\'{\i}nez}}\ and\ \bibinfo {author} {\bibfnamefont {J.~M.}\
  \bibnamefont {Hutson}},\ }\href@noop {} {\bibfield  {journal} {\bibinfo
  {journal} {Phys. Rev. A}\ }\textbf {\bibinfo {volume} {75}},\ \bibinfo
  {pages} {022702} (\bibinfo {year} {2007})}\BibitemShut {NoStop}%
\bibitem [{\citenamefont {Manolopoulos}(1986)}]{Manolopoulos:1986}%
  \BibitemOpen
  \bibfield  {author} {\bibinfo {author} {\bibfnamefont {D.~E.}\ \bibnamefont
  {Manolopoulos}},\ }\href@noop {} {\bibfield  {journal} {\bibinfo  {journal}
  {J. Chem. Phys.}\ }\textbf {\bibinfo {volume} {85}},\ \bibinfo {pages} {6425}
  (\bibinfo {year} {1986})}\BibitemShut {NoStop}%
\bibitem [{\citenamefont {Alexander}(1984)}]{Alexander:1984}%
  \BibitemOpen
  \bibfield  {author} {\bibinfo {author} {\bibfnamefont {M.~H.}\ \bibnamefont
  {Alexander}},\ }\href@noop {} {\bibfield  {journal} {\bibinfo  {journal} {J.
  Chem. Phys.}\ }\textbf {\bibinfo {volume} {81}},\ \bibinfo {pages} {4510}
  (\bibinfo {year} {1984})}\BibitemShut {NoStop}%
\bibitem [{\citenamefont {Hutson}(2007)}]{Hutson:res:2007}%
  \BibitemOpen
  \bibfield  {author} {\bibinfo {author} {\bibfnamefont {J.~M.}\ \bibnamefont
  {Hutson}},\ }\href@noop {} {\bibfield  {journal} {\bibinfo  {journal} {New J.
  Phys.}\ }\textbf {\bibinfo {volume} {9}},\ \bibinfo {pages} {152} (\bibinfo
  {year} {2007})}\BibitemShut {NoStop}%
\bibitem [{\citenamefont {Hutson}(1993)}]{Hutson:bound:1993}%
  \BibitemOpen
  \bibfield  {author} {\bibinfo {author} {\bibfnamefont {J.~M.}\ \bibnamefont
  {Hutson}},\ }\href@noop {} {\enquote {\bibinfo {title} {{BOUND} computer
  program, version 5},}\ }\bibinfo {howpublished} {distributed by Collaborative
  Computational Project No.\ 6 of the UK Engineering and Physical Sciences
  Research Council} (\bibinfo {year} {1993})\BibitemShut {NoStop}%
\bibitem [{\citenamefont {Hutson}(2011)}]{Hutson:field:2011}%
  \BibitemOpen
  \bibfield  {author} {\bibinfo {author} {\bibfnamefont {J.~M.}\ \bibnamefont
  {Hutson}},\ }\href@noop {} {\enquote {\bibinfo {title} {{FIELD} computer
  program, version 1},}\ } (\bibinfo {year} {2011})\BibitemShut {NoStop}%
\bibitem [{\citenamefont {Hutson}\ \emph {et~al.}(2008)\citenamefont {Hutson},
  \citenamefont {Tiesinga},\ and\ \citenamefont {Julienne}}]{Hutson:Cs2:2008}%
  \BibitemOpen
  \bibfield  {author} {\bibinfo {author} {\bibfnamefont {J.~M.}\ \bibnamefont
  {Hutson}}, \bibinfo {author} {\bibfnamefont {E.}~\bibnamefont {Tiesinga}}, \
  and\ \bibinfo {author} {\bibfnamefont {P.~S.}\ \bibnamefont {Julienne}},\
  }\href@noop {} {\bibfield  {journal} {\bibinfo  {journal} {Phys. Rev. A}\
  }\textbf {\bibinfo {volume} {78}},\ \bibinfo {pages} {052703} (\bibinfo
  {year} {2008})}\BibitemShut {NoStop}%
\bibitem [{\citenamefont {Strauss}\ \emph {et~al.}(2010)\citenamefont
  {Strauss}, \citenamefont {Takekoshi}, \citenamefont {Lang}, \citenamefont
  {Winkler}, \citenamefont {Grimm}, \citenamefont {Hecker~Denschlag},\ and\
  \citenamefont {Tiemann}}]{Strauss:2010}%
  \BibitemOpen
  \bibfield  {author} {\bibinfo {author} {\bibfnamefont {C.}~\bibnamefont
  {Strauss}}, \bibinfo {author} {\bibfnamefont {T.}~\bibnamefont {Takekoshi}},
  \bibinfo {author} {\bibfnamefont {F.}~\bibnamefont {Lang}}, \bibinfo {author}
  {\bibfnamefont {K.}~\bibnamefont {Winkler}}, \bibinfo {author} {\bibfnamefont
  {R.}~\bibnamefont {Grimm}}, \bibinfo {author} {\bibfnamefont
  {J.}~\bibnamefont {Hecker~Denschlag}}, \ and\ \bibinfo {author}
  {\bibfnamefont {E.}~\bibnamefont {Tiemann}},\ }\href {\doibase
  10.1103/PhysRevA.82.052514} {\bibfield  {journal} {\bibinfo  {journal} {Phys.
  Rev. A}\ }\textbf {\bibinfo {volume} {82}},\ \bibinfo {pages} {052514}
  (\bibinfo {year} {2010})}\BibitemShut {NoStop}%
\bibitem [{\citenamefont {Seto}\ \emph {et~al.}(2000)\citenamefont {Seto},
  \citenamefont {Le~Roy}, \citenamefont {Verg\`es},\ and\ \citenamefont
  {Amiot}}]{Seto:2000}%
  \BibitemOpen
  \bibfield  {author} {\bibinfo {author} {\bibfnamefont {J.~Y.}\ \bibnamefont
  {Seto}}, \bibinfo {author} {\bibfnamefont {R.~J.}\ \bibnamefont {Le~Roy}},
  \bibinfo {author} {\bibfnamefont {J.}~\bibnamefont {Verg\`es}}, \ and\
  \bibinfo {author} {\bibfnamefont {C.}~\bibnamefont {Amiot}},\ }\href@noop {}
  {\bibfield  {journal} {\bibinfo  {journal} {J. Chem. Phys}\ }\textbf
  {\bibinfo {volume} {113}},\ \bibinfo {pages} {3067} (\bibinfo {year}
  {2000})}\BibitemShut {NoStop}%
\bibitem [{\citenamefont {Moerdijk}\ \emph {et~al.}(1995)\citenamefont
  {Moerdijk}, \citenamefont {Verhaar},\ and\ \citenamefont
  {Axelsson}}]{Moerdijk:1995}%
  \BibitemOpen
  \bibfield  {author} {\bibinfo {author} {\bibfnamefont {A.~J.}\ \bibnamefont
  {Moerdijk}}, \bibinfo {author} {\bibfnamefont {B.~J.}\ \bibnamefont
  {Verhaar}}, \ and\ \bibinfo {author} {\bibfnamefont {A.}~\bibnamefont
  {Axelsson}},\ }\href@noop {} {\bibfield  {journal} {\bibinfo  {journal}
  {Phys. Rev. A}\ }\textbf {\bibinfo {volume} {51}},\ \bibinfo {pages} {4852}
  (\bibinfo {year} {1995})}\BibitemShut {NoStop}%
\bibitem [{\citenamefont {Hutson}\ \emph {et~al.}(2009)\citenamefont {Hutson},
  \citenamefont {Beyene},\ and\ \citenamefont
  {Gonz\'{a}lez-Mart\'{\i}nez}}]{Hutson:HeO2:2009}%
  \BibitemOpen
  \bibfield  {author} {\bibinfo {author} {\bibfnamefont {J.~M.}\ \bibnamefont
  {Hutson}}, \bibinfo {author} {\bibfnamefont {M.}~\bibnamefont {Beyene}}, \
  and\ \bibinfo {author} {\bibfnamefont {M.~L.}\ \bibnamefont
  {Gonz\'{a}lez-Mart\'{\i}nez}},\ }\href@noop {} {\bibfield  {journal}
  {\bibinfo  {journal} {Phys. Rev. Lett.}\ }\textbf {\bibinfo {volume} {103}},\
  \bibinfo {pages} {163201} (\bibinfo {year} {2009})}\BibitemShut {NoStop}%
\bibitem [{\citenamefont {McCarron}\ \emph {et~al.}(2011)\citenamefont
  {McCarron}, \citenamefont {Cho}, \citenamefont {Jenkin}, \citenamefont
  {K{\"{o}}ppinger},\ and\ \citenamefont {Cornish}}]{McCarron2011}%
  \BibitemOpen
  \bibfield  {author} {\bibinfo {author} {\bibfnamefont {D.~J.}\ \bibnamefont
  {McCarron}}, \bibinfo {author} {\bibfnamefont {H.~W.}\ \bibnamefont {Cho}},
  \bibinfo {author} {\bibfnamefont {D.~L.}\ \bibnamefont {Jenkin}}, \bibinfo
  {author} {\bibfnamefont {M.~P.}\ \bibnamefont {K{\"{o}}ppinger}}, \ and\
  \bibinfo {author} {\bibfnamefont {S.~L.}\ \bibnamefont {Cornish}},\
  }\href@noop {} {\bibfield  {journal} {\bibinfo  {journal} {Phys. Rev. A}\
  }\textbf {\bibinfo {volume} {84}},\ \bibinfo {pages} {011603} (\bibinfo
  {year} {2011})}\BibitemShut {NoStop}%
\bibitem [{\citenamefont {Cho}\ \emph {et~al.}(2011)\citenamefont {Cho},
  \citenamefont {McCarron}, \citenamefont {Jenkin}, \citenamefont
  {K{\"{o}}ppinger},\ and\ \citenamefont {Cornish}}]{Cho2011}%
  \BibitemOpen
  \bibfield  {author} {\bibinfo {author} {\bibfnamefont {H.~W.}\ \bibnamefont
  {Cho}}, \bibinfo {author} {\bibfnamefont {D.~J.}\ \bibnamefont {McCarron}},
  \bibinfo {author} {\bibfnamefont {D.~L.}\ \bibnamefont {Jenkin}}, \bibinfo
  {author} {\bibfnamefont {M.~P.}\ \bibnamefont {K{\"{o}}ppinger}}, \ and\
  \bibinfo {author} {\bibfnamefont {S.~L.}\ \bibnamefont {Cornish}},\
  }\href@noop {} {\bibfield  {journal} {\bibinfo  {journal} {Eur. Phys. J. D}\
  }\textbf {\bibinfo {volume} {65}},\ \bibinfo {pages} {125} (\bibinfo {year}
  {2011})}\BibitemShut {NoStop}%
\bibitem [{\citenamefont {Lin}\ \emph {et~al.}(2009)\citenamefont {Lin},
  \citenamefont {Perry}, \citenamefont {Compton}, \citenamefont {Spielman},\
  and\ \citenamefont {Porto}}]{Lin2009}%
  \BibitemOpen
  \bibfield  {author} {\bibinfo {author} {\bibfnamefont {Y.-J.}\ \bibnamefont
  {Lin}}, \bibinfo {author} {\bibfnamefont {A.~R.}\ \bibnamefont {Perry}},
  \bibinfo {author} {\bibfnamefont {R.~L.}\ \bibnamefont {Compton}}, \bibinfo
  {author} {\bibfnamefont {I.~B.}\ \bibnamefont {Spielman}}, \ and\ \bibinfo
  {author} {\bibfnamefont {J.~V.}\ \bibnamefont {Porto}},\ }\href@noop {}
  {\bibfield  {journal} {\bibinfo  {journal} {Phys. Rev. A}\ }\textbf {\bibinfo
  {volume} {79}},\ \bibinfo {pages} {063631} (\bibinfo {year}
  {2009})}\BibitemShut {NoStop}%
\bibitem [{\citenamefont {Jenkin}\ \emph {et~al.}(2011)\citenamefont {Jenkin},
  \citenamefont {McCarron}, \citenamefont {K{\"{o}}ppinger}, \citenamefont
  {Cho}, \citenamefont {Hopkins},\ and\ \citenamefont {Cornish}}]{Jenkin2011}%
  \BibitemOpen
  \bibfield  {author} {\bibinfo {author} {\bibfnamefont {D.~L.}\ \bibnamefont
  {Jenkin}}, \bibinfo {author} {\bibfnamefont {D.~J.}\ \bibnamefont
  {McCarron}}, \bibinfo {author} {\bibfnamefont {M.~P.}\ \bibnamefont
  {K{\"{o}}ppinger}}, \bibinfo {author} {\bibfnamefont {H.~W.}\ \bibnamefont
  {Cho}}, \bibinfo {author} {\bibfnamefont {S.~A.}\ \bibnamefont {Hopkins}}, \
  and\ \bibinfo {author} {\bibfnamefont {S.~L.}\ \bibnamefont {Cornish}},\
  }\href@noop {} {\bibfield  {journal} {\bibinfo  {journal} {Eur. Phys. J. D}\
  }\textbf {\bibinfo {volume} {65}},\ \bibinfo {pages} {11} (\bibinfo {year}
  {2011})}\BibitemShut {NoStop}%
\bibitem [{\citenamefont {Altin}\ \emph {et~al.}(2010)\citenamefont {Altin},
  \citenamefont {Robins}, \citenamefont {Poldy}, \citenamefont {Debs},
  \citenamefont {D\"oring}, \citenamefont {Figl},\ and\ \citenamefont
  {Close}}]{Altin:2010b}%
  \BibitemOpen
  \bibfield  {author} {\bibinfo {author} {\bibfnamefont {P.~A.}\ \bibnamefont
  {Altin}}, \bibinfo {author} {\bibfnamefont {N.~P.}\ \bibnamefont {Robins}},
  \bibinfo {author} {\bibfnamefont {R.}~\bibnamefont {Poldy}}, \bibinfo
  {author} {\bibfnamefont {J.~E.}\ \bibnamefont {Debs}}, \bibinfo {author}
  {\bibfnamefont {D.}~\bibnamefont {D\"oring}}, \bibinfo {author}
  {\bibfnamefont {C.}~\bibnamefont {Figl}}, \ and\ \bibinfo {author}
  {\bibfnamefont {J.~D.}\ \bibnamefont {Close}},\ }\href {\doibase
  10.1103/PhysRevA.81.012713} {\bibfield  {journal} {\bibinfo  {journal} {Phys.
  Rev. A}\ }\textbf {\bibinfo {volume} {81}},\ \bibinfo {pages} {012713}
  (\bibinfo {year} {2010})}\BibitemShut {NoStop}%
\bibitem [{\citenamefont {Berninger}\ \emph {et~al.}(2011)\citenamefont
  {Berninger}, \citenamefont {Zenesini}, \citenamefont {Huang}, \citenamefont
  {Harm}, \citenamefont {N\"agerl}, \citenamefont {Ferlaino}, \citenamefont
  {Grimm}, \citenamefont {Julienne},\ and\ \citenamefont
  {Hutson}}]{Berninger:Efimov:2011}%
  \BibitemOpen
  \bibfield  {author} {\bibinfo {author} {\bibfnamefont {M.}~\bibnamefont
  {Berninger}}, \bibinfo {author} {\bibfnamefont {A.}~\bibnamefont {Zenesini}},
  \bibinfo {author} {\bibfnamefont {B.}~\bibnamefont {Huang}}, \bibinfo
  {author} {\bibfnamefont {W.}~\bibnamefont {Harm}}, \bibinfo {author}
  {\bibfnamefont {H.-C.}\ \bibnamefont {N\"agerl}}, \bibinfo {author}
  {\bibfnamefont {F.}~\bibnamefont {Ferlaino}}, \bibinfo {author}
  {\bibfnamefont {R.}~\bibnamefont {Grimm}}, \bibinfo {author} {\bibfnamefont
  {P.~S.}\ \bibnamefont {Julienne}}, \ and\ \bibinfo {author} {\bibfnamefont
  {J.~M.}\ \bibnamefont {Hutson}},\ }\href@noop {} {\bibfield  {journal}
  {\bibinfo  {journal} {Phys. Rev. Lett.}\ }\textbf {\bibinfo {volume} {107}},\
  \bibinfo {pages} {120401} (\bibinfo {year} {2011})}\BibitemShut {NoStop}%
\bibitem [{\citenamefont {Ferlaino}\ \emph {et~al.}(2008)\citenamefont
  {Ferlaino}, \citenamefont {Knoop}, \citenamefont {Mark}, \citenamefont
  {Berninger}, \citenamefont {Sch\"obel}, \citenamefont {N\"agerl},\ and\
  \citenamefont {Grimm}}]{Ferlaino:halo:2008}%
  \BibitemOpen
  \bibfield  {author} {\bibinfo {author} {\bibfnamefont {F.}~\bibnamefont
  {Ferlaino}}, \bibinfo {author} {\bibfnamefont {S.}~\bibnamefont {Knoop}},
  \bibinfo {author} {\bibfnamefont {M.}~\bibnamefont {Mark}}, \bibinfo {author}
  {\bibfnamefont {M.}~\bibnamefont {Berninger}}, \bibinfo {author}
  {\bibfnamefont {H.}~\bibnamefont {Sch\"obel}}, \bibinfo {author}
  {\bibfnamefont {H.-C.}\ \bibnamefont {N\"agerl}}, \ and\ \bibinfo {author}
  {\bibfnamefont {R.}~\bibnamefont {Grimm}},\ }\href {\doibase
  10.1103/PhysRevLett.101.023201} {\bibfield  {journal} {\bibinfo  {journal}
  {Phys. Rev. Lett.}\ }\textbf {\bibinfo {volume} {101}},\ \bibinfo {pages}
  {023201} (\bibinfo {year} {2008})}\BibitemShut {NoStop}%
\bibitem [{\citenamefont {Weber}\ \emph {et~al.}(2003)\citenamefont {Weber},
  \citenamefont {Herbig}, \citenamefont {Mark}, \citenamefont {Nägerl},\ and\
  \citenamefont {Grimm}}]{Weber:2003}%
  \BibitemOpen
  \bibfield  {author} {\bibinfo {author} {\bibfnamefont {T.}~\bibnamefont
  {Weber}}, \bibinfo {author} {\bibfnamefont {J.}~\bibnamefont {Herbig}},
  \bibinfo {author} {\bibfnamefont {M.}~\bibnamefont {Mark}}, \bibinfo {author}
  {\bibfnamefont {H.-C.}\ \bibnamefont {Nägerl}}, \ and\ \bibinfo {author}
  {\bibfnamefont {R.}~\bibnamefont {Grimm}},\ }\href {\doibase
  10.1126/science.1079699} {\bibfield  {journal} {\bibinfo  {journal}
  {Science}\ }\textbf {\bibinfo {volume} {299}},\ \bibinfo {pages} {232}
  (\bibinfo {year} {2003})}\BibitemShut {NoStop}%
\bibitem [{\citenamefont {Marchant}\ \emph {et~al.}(2012)\citenamefont
  {Marchant}, \citenamefont {H\"andel}, \citenamefont {Hopkins}, \citenamefont
  {Wiles},\ and\ \citenamefont {Cornish}}]{Marchant:2012}%
  \BibitemOpen
  \bibfield  {author} {\bibinfo {author} {\bibfnamefont {A.~L.}\ \bibnamefont
  {Marchant}}, \bibinfo {author} {\bibfnamefont {S.}~\bibnamefont {H\"andel}},
  \bibinfo {author} {\bibfnamefont {S.~A.}\ \bibnamefont {Hopkins}}, \bibinfo
  {author} {\bibfnamefont {T.~P.}\ \bibnamefont {Wiles}}, \ and\ \bibinfo
  {author} {\bibfnamefont {S.~L.}\ \bibnamefont {Cornish}},\ }\href {\doibase
  10.1103/PhysRevA.85.053647} {\bibfield  {journal} {\bibinfo  {journal} {Phys.
  Rev. A}\ }\textbf {\bibinfo {volume} {85}},\ \bibinfo {pages} {053647}
  (\bibinfo {year} {2012})}\BibitemShut {NoStop}%
\bibitem [{\citenamefont {Chin}\ \emph {et~al.}(2004)\citenamefont {Chin},
  \citenamefont {Vuleti\'c}, \citenamefont {Kerman}, \citenamefont {Chu},
  \citenamefont {Tiesinga}, \citenamefont {Leo},\ and\ \citenamefont
  {Williams}}]{Chin:cs2-fesh:2004}%
  \BibitemOpen
  \bibfield  {author} {\bibinfo {author} {\bibfnamefont {C.}~\bibnamefont
  {Chin}}, \bibinfo {author} {\bibfnamefont {V.}~\bibnamefont {Vuleti\'c}},
  \bibinfo {author} {\bibfnamefont {A.~J.}\ \bibnamefont {Kerman}}, \bibinfo
  {author} {\bibfnamefont {S.}~\bibnamefont {Chu}}, \bibinfo {author}
  {\bibfnamefont {E.}~\bibnamefont {Tiesinga}}, \bibinfo {author}
  {\bibfnamefont {P.~J.}\ \bibnamefont {Leo}}, \ and\ \bibinfo {author}
  {\bibfnamefont {C.~J.}\ \bibnamefont {Williams}},\ }\href@noop {} {\bibfield
  {journal} {\bibinfo  {journal} {Phys. Rev. A}\ }\textbf {\bibinfo {volume}
  {70}},\ \bibinfo {pages} {032701} (\bibinfo {year} {2004})}\BibitemShut
  {NoStop}%
\bibitem [{\citenamefont {Berninger}\ \emph {et~al.}(2012)\citenamefont
  {Berninger}, \citenamefont {Zenesini}, \citenamefont {Huang}, \citenamefont
  {Harm}, \citenamefont {N\"agerl}, \citenamefont {Ferlaino}, \citenamefont
  {Grimm}, \citenamefont {Julienne},\ and\ \citenamefont
  {Hutson}}]{Berninger:Cs2:2013}%
  \BibitemOpen
  \bibfield  {author} {\bibinfo {author} {\bibfnamefont {M.}~\bibnamefont
  {Berninger}}, \bibinfo {author} {\bibfnamefont {A.}~\bibnamefont {Zenesini}},
  \bibinfo {author} {\bibfnamefont {B.}~\bibnamefont {Huang}}, \bibinfo
  {author} {\bibfnamefont {W.}~\bibnamefont {Harm}}, \bibinfo {author}
  {\bibfnamefont {H.-C.}\ \bibnamefont {N\"agerl}}, \bibinfo {author}
  {\bibfnamefont {F.}~\bibnamefont {Ferlaino}}, \bibinfo {author}
  {\bibfnamefont {R.}~\bibnamefont {Grimm}}, \bibinfo {author} {\bibfnamefont
  {P.~S.}\ \bibnamefont {Julienne}}, \ and\ \bibinfo {author} {\bibfnamefont
  {J.~M.}\ \bibnamefont {Hutson}},\ }\href@noop {} {\bibfield  {journal}
  {\bibinfo  {journal} {arXiv}\ ,\ \bibinfo {pages} {2012:12xx}} (\bibinfo
  {year} {2012})}\BibitemShut {NoStop}%
\end{thebibliography}%

\end{document}